\documentclass[12pt,a4]{article}
\usepackage{amsthm,amsmath,latexsym,amssymb,amsfonts,amscd}
\usepackage{graphics,lscape,fancyhdr,array,stmaryrd,euscript}
\usepackage{ifthen,empheq,slashed}
\numberwithin{equation}{section}
\usepackage{setspace}

\usepackage{hyperref}
\usepackage{cite}

\usepackage[small,nohug,heads=littlevee,l>=20pt]{diagrams}

\begin{document}
\hfill
\begin{flushright}
    {LMU-ASC 52/17}
\end{flushright}
\vskip 0.02\textheight
\begin{center}

{\Large\bfseries Elementary particles with continuous spin
\vspace{0.4cm}
} \\

\vskip 0.04\textheight

Xavier \textsc{Bekaert},${}^{1,2}$ Evgeny \textsc{Skvortsov},${}^{3,4}$

\vskip 0.04\textheight

{\em ${}^1$Laboratoire de Math\'ematiques et Physique Th\'eorique\\
Unit\'e Mixte de Recherche $7350$ du CNRS\\
F\'ed\'eration de Recherche $2964$ Denis Poisson\\
Universit\'e Fran\c{c}ois Rabelais, Parc de Grandmont\\
37200 Tours, France}\\
\vspace*{5pt}
{\em ${}^2$B.W. Lee Center for Fields, Gravity and Strings\\ 
Institute for Basic Science\\ 
Daejeon, South Korea}\\
\vspace*{5pt}
{\em ${}^{3}$ Arnold Sommerfeld Center for Theoretical Physics\\
Ludwig-Maximilians University Munich\\
Theresienstr. 37, D-80333 Munich, Germany}\\
\vspace*{5pt}
{\em ${}^{4}$ Lebedev Institute of Physics, \\
Leninsky ave. 53, 119991 Moscow, Russia}

\vskip 0.02\textheight

{\bf Abstract }

\end{center}
\begin{quotation}
Classical results and recent developments on the theoretical description of elementary particles with ``continuous'' spin are reviewed. At free level, these fields are described by unitary irreducible representations of the isometry group (either Poincar\'e or anti de Sitter group) with an infinite number of physical degrees of freedom per spacetime point. Their basic group-theoretical and field-theoretical descriptions are reviewed in some details. We mention a list of open issues which are crucial to address for assessing their physical status and potential relevance.
\end{quotation}

\newpage

\tableofcontents
\newpage
\section{Introduction}

Wigner taught \cite{Wigner:1939cj} us that free elementary particles propagating on Minkowski spacetime are in one-to-one correspondence with unitary irreducible representations (UIRs) of the Poincar\'e group.\footnote{Apart from Wigner's seminal paper \cite{Wigner:1939cj} on the subject, another classical introduction to this subject is the short review on this topic inside his seminal work \cite{Bargmann:1948ck} with Bargmann on relativistic wave equations.} This motivated his classification \cite{Wigner:1939cj} of all such representations in the case of  spacetime dimension four. In this section, we briefly review the main topics and results to be covered in the body of the paper.

\subsection{Definition and exotic properties}

Among the massless representations, the so-called ``continuous-spin'' representations are actually the generic ones. In fact, the eigenvalue of the quadratic Casimir operator (= the square of the momentum $P_\mu$) vanishes for these representations (i.e. they are massless), 
\begin{equation}
{\cal C}_2\Big(\mathfrak{iso}(3,1)\Big):=-P^2=0\,,
\label{C2}
\end{equation}
but \textit{not} the quartic Casimir operator (= the square of the Pauli-Lubanski vector $W_\mu$) which is a positive real number, 
\begin{equation}
{\cal C}_4\Big(\mathfrak{iso}(3,1)\Big):=W^2=\mu^2>0\,,
\label{C4}
\end{equation}
where $W_\mu:=\frac12\epsilon_{\mu\nu\rho\sigma}P^\nu J^{\nu\rho}$\,, with $\mathfrak{so}(3,1)$\,=\,span$\{J_{\mu\nu}\}$.
Notice that we work in the ``mostly plus'' signature.

Although they appear in a natural way and are somewhat generic from a mathematical point of view (as UIRs), elementary particles described by continuous-spin representations are usually discarded by high-energy theoretical physicists on the basis of two exotic features:
\begin{enumerate}
 \item \textit{They are characterized by a continuous parameter with the dimension of a mass, although they are massless.}
 \item \textit{They have infinitely many degrees of freedom per spacetime point.}
\end{enumerate}

The first exotic property is the origin of the unfortunate terminology ``continuous-spin''. These UIRs are indeed characterized by the continuous parameter $\mu>0$ but it is somewhat misleading to interpret it as related to spin. Actually the ``spin'' of these particles is by no means continuous, in contrast with anyons in three dimensions. Rather, the helicity eigenvalues are discrete: either all integers, or all half-integers, in four-dimensional spacetime.

The meaning of the second property is that the helicities in their spectrum are unbounded.  More precisely, these particles are described by the countably infinite tower of all helicity states (either all integer or all half-integer helicities) mixing under Lorentz boosts. Since one definition of the ``spin'' is as the bound on the (absolute value of the) helicity eigenvalues, Wigner later \cite{Wigner:1963in} proposed the alternative (presumably better) terminology ``infinite spin'' to refer to these exotic representations. Unfortunately, terminological habits are difficult to change so we will mostly stick to the standard terminology in this review paper.

\subsection{Speculations}

The infinite number of degrees of freedom (per spacetime point) was the main reason of Wigner's rejection of the continuous-spin representation. He claimed that this property implies that the heat capacity of a gas of such particles is infinite \cite{Wigner:1963in}.

However, from the point of view of higher-spin gravity the two exotic properties of continuous-spin particles can be turned upside-down as two positive qualitative features to expect for a candidate theory of interacting massless particles of higher helicities in a flat spacetime.
Firstly, an infinite spectrum of helicities is a standard feature of higher-spin theories in dimension four (and higher). More importantly, the spectrum of helicities of a continuous-spin particle coincides with the one in higher-spin gravity.\footnote{More precisely, the helicity spectra coincide for the bosonic case in $D\geqslant4$ and for its supersymmetric extension in $D=4$.} Secondly, higher-spin vertices are typically higher-derivative, thus a dimensionful parameter for weighting them is a necessary feature of any interacting theory. 
In fact, it has been suggested \cite{Bekaert:2005in} that continuous-spin gauge fields might be able to circumvent the no-go theorems\footnote{See e.g. \cite{Bekaert:2010hw,Rahman:2013sta} for some introductory reviews of these no-go theorems.} preventing the existence of interacting particles of spin greater than two in flat spacetime via a mechanism similar to what happens \cite{Fradkin:1986qy} in the presence of a cosmological constant (which plays the role of the dimensionful parameter). Accordingly, one may speculate that they might provide a subtle flat spacetime analogue of higher-spin gravity, since both spectra coincide at free level.

Unfortunately, although this exotic representation is known for many decades, the literature and positive results on continuous-spin particles are scarce because they are usually discarded without serious scrutiny.
However, Schuster and Toro recently proposed a class of soft factors for these massless particles, from which they argued that the phenomenology of continuous-spin particles might be better behaved than expected, in the sense that \cite{Schuster:2013pxj,Schuster:2013vpr}:
\begin{itemize}
	\item These particles might circumvent Weinberg's no-go theorem  \cite{Weinberg:1964ew} on long-range interactions mediated by massless particles of helicity higher than two in flat spacetime.
	\item At energies higher than the characteristic mass parameter alluded above, they may experience ``helicity correspondence'' in that they effectively behave like massless particles with helicities not higher than two.
\end{itemize}
These expectations are suggestive but these authors had to postulate a suitable class of form factors which, to be confirmed, should be derived from first principles. 

Due to these various motivations, these exotic representations have recently attracted some attention, though many open problems within reach remain to be explored in order to clarify the status of the previous suggestive scenarios.

\subsection{State of the art and main challenges}	

In order to investigate the properties of these exotic particles without any prejudice and reach definite conclusions about their phenomenological viability, one should start by developing further their field-theoretic description. To be more specific, one should achieve on a first-principle basis the following tasks: 
\begin{itemize}
	\item \textbf{Kinematics:} 
	present covariant descriptions, on-shell (linear wave equations) and off-shell (quadratic Lagrangians).
	\item \textbf{Dynamics:} 
	classify their consistent interactions, off-shell (vertices) and on-shell (scattering amplitudes).
\end{itemize} 
Although their dynamics is a rather challenging problem which has not been explored yet, some progress has been made on the front of kinematics during the last decade.\footnote{We will not discuss here the purely quantum field-theoretical approach to continuous-spin particles \cite{Mund:2004sy,Schroer:2015rct}. The results there show that, under some general hypotheses, quantum fields transforming as continuous spin representations cannot have pointlike localization and are, at best, localized on semi-infinite spacelike strings.}

While covariant wave equations describing a continuous-spin particle date back to Wigner himself \cite{Wigner:1947}, the inverse variational problem\footnote{The inverse variational problem is usually referred to as the ``Fierz-Pauli programme'' in the higher-spin literature.} (i.e. obtaining an action from which they derive as Euler-Lagrange equations) remained an important open problem for quite a long time. A first step in this direction was made by obtaining a gauge formulation  \cite{Bekaert:2005in} for these fields in terms of a deformation of the Fronsdal equation \cite{Fronsdal:1978rb,Fang:1978wz}, which was shown to be equivalent to Wigner's original system of equations \cite{Wigner:1947}. The equation of Fronsdal describes a massless particle with discrete/finite spin (i.e. a helicity UIR of the Poincar\'e group) and is associated to an action principle \cite{Fronsdal:1978rb,Fang:1978wz} so this first step was encouraging. Recently, action principles for continuous-spin gauge fields have been proposed in the bosonic \cite{Schuster:2013pta} and fermionic \cite{Najafizadeh:2015uxa} cases, which are analogous to the action proposed by Segal \cite{Segal:2001qq} for bosonic higher-spin gauge fields on (anti) de Sitter spacetime (see also  \cite{Rivelles:2014fsa} for some analyses of the bosonic action  \cite{Schuster:2013pta}). 
A remarkable new step forward was performed by Metsaev who proposed \cite{Metsaev:2016lhs,Metsaev:2017ytk} action principles for continuous/infinite spin particles, either on Minkowski or on (anti) de Sitter spacetimes. These actions have the virtue of being deformations of the action principles of Fronsdal \cite{Fronsdal:1978rb,Fang:1978wz,Fronsdal:1978vb,Fang:1979hq} to which they reduce in the limit $\mu\to 0$. It turned out that the Metsaev action is directly related to the action for massive higher-spin fields proposed by Zinoviev long ago \cite{Zinoviev:2001dt}. Also, an unfolded formulation for continuous-spin fields was implicitly constructed \cite{Ponomarev:2010st} in the course of constructing the unfolded formulation of massive higher-spin fields. Lastly, wave equations for continuous-spin particles were derived from first-quantization of particle/string models \cite{BRSTpack}.

Although the first task (kinematics) in the list has essentially been completed since Lagrangians have been proposed for continuous-spin gauge fields, one should emphasize that an important consistency check remains to be done: by extracting the corresponding propagator and computing the corresponding current exchange, one should check that only physical degrees of freedom propagate. Indeed, as exhibited by the example of higher spins, some subtleties may arise at the level of the propagator \cite{Francia:2007qt}.

Anyway, much more challenging is the second task (dynamics).
A preliminary investigation would be the classification of the consistent cubic vertices for continuous-spin particles, either self-interacting or interacting with lower-spin matter (such as scalar fields, gauge vectors, or gravitons). Even a negative answer (the absence of consistent interactions) would be an important result since it would provide a no-go theorem against the physical relevance of these representations. On the contrary, a positive result (the existence of cubic self-interactions) would be a very strong indication that continuous-spin particles might provide the proper flat spacetime analogue of higher-spin gravity.

\subsection{Plan of the paper}

In section \ref{Wclass}, the continuous-spin particles on flat spacetime are introduced from a group-theoretical point of view via their classification, either following the method of induced representations or according to the eigenvalues of Casimir operators.
In section \ref{WEqs}, we provide a review of covariant equations describing either massive (or massless) representations of discrete/finite spin or massless representations of continuous/infinite spin, as well as the relations between them.
In section \ref{Action}, we briefly discuss action principles for bosonic (higher-spin or continuous-spin) gauge fields on flat spacetime. In section \ref{AdS}, we chart the landscape of unitary continuous-spin fields on anti de Sitter spacetime, found by Metsaev in his exploration of action principles on constant-curvature spacetimes, and the relation to earlier constructions by Zinoviev is clarified. Also in section \ref{AdS}, we discuss the one-loop partition functions. We conclude with a list of open problems in section \ref{conclu}.

\section{Wigner classification}\label{Wclass}

Let us start by reviewing the representation theory of the Poincar\'e group.\footnote{See e.g. \cite{Bekaert:2006py} for a pedagogical introduction to the classification in any dimension.}  Actually, we will restrict our attention to the Lie algebra for simplicity.

\subsection{Method of induced representations for the Poincar\'e algebra}

The Poincar\'e algebra $\mathfrak{iso}(D-1,1)={\mathbb R}^D\niplus\mathfrak{so}(D-1,1)$ is the semidirect sum of the Abelian algebra ${\mathbb R}^D$ and the semisimple Lie algebra $\mathfrak{so}(D-1,1)$, therefore
Wigner's method of induced representation applies and amounts to the following steps:
	\begin{enumerate}
		\item Consider the UIRs of the Abelian subalgebra ${\mathbb R}^D$: they are labelled by real eigenvalues (unitary \& irreducible) of the generators $\hat{P}_\mu$, i.e. by the momentum $p_\mu$.
		\item Identify the orbit and stabilizer of these eigenvalues: they are, respectively, the ``mass shell'' and ``little group'' in physicist language.
		\item Induce the UIR of the full algebra from a UIR of the stability subalgebra:
this latter representation spans the ``spinning'' degrees of freedom (or ``physical components'') of the corresponding field.
	\end{enumerate}
The various possibilities are summarized in Table \ref{Table1}. 

\begin{table}
\caption{Classification of the UIRs of the Poincar\'e algebra  $\mathfrak{iso}(D-1,1)$}
\begin{center}
{\begin{tabular}{|c|c|c|c|}
 \hline
  UIR & Orbit & Stability & \# of components \\\hline\hline
  \small{Massive} & \tiny{2-sheeted hyperboloid} $p^2=-m^2$ & $\mathfrak{so}(D-1)$ & \small{finite}\\\hline
  \small{Massless} & \tiny{light-cone} $p^2=0$ & $\mathfrak{iso}(D-2)$ & \small{finite or $\infty$} \\\hline
  \small{Tachyonic} & \tiny{1-sheeted hyperboloid} $p^2=+m^2$ & $\mathfrak{so}(D-2,1)$ & \small{1 or $\infty$} \\\hline
  \small{Zero-momentum} & \tiny{origin} $p_\mu=0$ & $\mathfrak{so}(D-1,1)$ & \small{unfaithful irrep} \\\hline
\end{tabular}}
\end{center}
\label{Table1}
\end{table}

The number of physical degrees of freedom per spacetime point (also called the number of independent physical components) is the dimension of the UIR of the little group. A trivial representation of the little group corresponds to a scalar field. In the other cases, the number of physical components is finite if and only if the little group is compact. In particular, the massive representations always have a finite number of components. In contrast, the only tachyonic representation which has a finite number of components is a scalar tachyonic field. 

\subsection{Massless representations}

Consider a massless  particle in $D$-dimensional Minkowski spacetime ${\mathbb R}^{D-1,1}$, with light-like momentum $p_\mu$ ($\mu=0,1,2,\cdots,D-1$). 
A spacelike plane orthogonal to this light-like momentum will be called a ``transverse plane'' ${\mathbb R}^{D-2}\subset{\mathbb R}^{D-1,1}$. The generators of the Lorentz algebra $\mathfrak{so}(D-1,1)$ are denoted $\hat{J}_{\mu\nu}$ and obey the commutation relations
\begin{equation}
\left[\hat{J}_{\mu\nu},\hat{J}_{\rho\sigma}\right]=
 i\,\big(g_{\nu\rho}\hat{J}_{\mu\sigma}
 -g_{\mu\rho}\hat{J}_{\nu\sigma}-g_{\nu\sigma}\hat{J}_{\mu\rho}+g_{\mu\sigma}\hat{J}_{\nu\rho}\big)\,,
\end{equation}
where $g_{\mu\nu}$ is the Minkowski metric (in the ``mostly plus'' signature).

To discuss massless particles, it turns out to be convenient to use light-cone coordinates 
$x^{\pm}={1 \over \sqrt{2}}\big(x^0\pm x^{D-1}\big)$
adapted to the momentum, i.e. the latter has zero components except for $p^+=-p_-\,$. The Minkowski metric reads
\begin{equation}
ds^2\,=\,-\,2\,dx^+dx^-\,+\,dx^idx_i\,,\qquad (i=1,2,\cdots,D-2)\,,
\end{equation}
where $x^i$ are Cartesian coordinates on the transverse plane.

The massless little group $ISO(D-2)$ leaving the momentum invariant is formed of:
\begin{itemize}
	\item rotations of the transverse plane, generated by $$\hat{J}_{ij}\qquad\qquad (i,j=1,2,\cdots,D-2)$$
	\item transverse null boosts, generated by $$\hat{\pi}_i:=\hat{J}_{+i}={1 \over \sqrt{2}}\big(\hat{J}_{0i}+ \hat{J}_{D-1\,i}\big)\qquad (i=1,2,\cdots,D-2)\,.$$ 
\end{itemize}

The generators span a Lie algebra isomorphic to the Euclidean algebra $\mathfrak{iso}(D-2)$ of the transverse plane
\begin{equation}
 \left[\hat{J}_{ij},\hat{J}_{kl}\right]=
 i\,(\delta_{jk}\hat{J}_{il}
 -\delta_{ik}\hat{J}_{jl}-\delta_{jl}\hat{J}_{ik}+\delta_{il}\hat{J}_{jk})\,,
\end{equation}
\begin{equation}
 \left[\hat{\pi}_i,\hat{J}_{kl}\right]=i\,(\delta_{ik}\hat{\pi}_l-\delta_{il}\hat{\pi}_k)\,,
\end{equation}
\begin{equation}
 \left[\hat{\pi}_i,\hat{\pi}_j\right] = 0\,.
\end{equation}
Therefore, the problem of classifying the massless UIRs of the Poincar\'e algebra $\mathfrak{iso}(D-1,1)$ reduces to the classification of the UIRs of the Euclidean algebra $\mathfrak{iso}(D-2)$. 

The Euclidean algebra $\mathfrak{iso}(D-2)$ of the transverse plane is noncompact and all its nontrivial faithful UIRs are infinite-dimensional.
Thus the massless representations either (i) have a finite number of components corresponding to trivial or unfaithful representations of $\mathfrak{iso}(D-2)$, or (ii) have an infinite number of components corresponding to faithful representations of $\mathfrak{iso}(D-2)$:
\begin{itemize}
	\item[(i)] \textbf{Discrete/finite spin representations:} The first case of massless representations are the ``helicity'' (also called ``discrete/finite-spin'') representations. They arise from a trivial representation of the transverse ``translation'' subalgebra ${\mathbb R}^{D-2}\subset\mathfrak{iso}(D-2)$, thus effectively, they are induced from a UIR of the transverse rotation subalgebra
$\mathfrak{so}(D-2)\subset\mathfrak{iso}(D-2)$. The latter algebra is compact and semisimple, therefore it only has finite-dimensional UIRs.
The corresponding fields have indeed a finite number of components (``finite spin'')
	\item[(ii)] \textbf{Continuous/infinite spin representations:} The second case of massless representations arise from nontrivial representations of the transverse ``translation'' subalgebra ${\mathbb R}^{D-2}\subset\mathfrak{iso}(D-2)$, hence they are induced from infinite-dimensional UIRs of the little group $ISO(D-2)$.
\end{itemize}

In fact, let us consider more closely the representation of the massless little group $ISO(D-2)$ by applying the method of induced representations to the Euclidean group $ISO(d)$ on its own.
The Euclidean algebra $\mathfrak{iso}(d)={\mathbb R}^{d}\niplus\mathfrak{so}(d)$ is the semidirect sum of the Abelian algebra ${\mathbb R}^{d}$ and the semisimple Lie algebra $\mathfrak{so}(d)$. Wigner's {method of induced representation}, applied to this case, goes as follows:
	\begin{enumerate}
		\item Consider the UIRs of the Abelian subalgebra ${\mathbb R}^{d}$: they are labelled by real eigenvalues of the generators $\hat{\pi}_i$ ($i=1,2,\cdots,d$). This defines a vector $w_i$ of the plane.
		\item Identify the orbit and stabilizer of these eigenvalues.
		\item Induce the UIR of the Euclidean algebra from an UIR of the stability subalgebra.
\end{enumerate}
The result is summarized in Table \ref{Table2}.

\begin{table}
\caption{Classification of the UIRs of the Euclidean algebra  $\mathfrak{iso}(d)$}
\begin{center}
{\begin{tabular}{|c|c|c|c|c|}
 \hline
  UIR & Dimension & Orbit & Stability & Example\\\hline\hline
  \small{Unfaithful} & \small{finite} & \tiny{origin} $w_i=0$ & $\mathfrak{so}(d)$ & \tiny{Spherical harmonics}  \\\hline
  \small{Faithful} & \small{infinite} & \tiny{sphere} $w^2=\mu^2$ & $\mathfrak{so}(d-1)$ & \tiny{Solutions of Helmholtz eq} 
	\\\hline
\end{tabular}
\label{Table2}
}
\end{center}
\end{table}

Particularizing this result to $d=D-2$ and looking for the representations of the Poincar\'e algebra $\mathfrak{iso}(D-1,1)$ induced from the Euclidean algebra $\mathfrak{iso}(D-2)$, one finds the result charted in Table \ref{Table3}.

\begin{table}
\caption{Classification of the massless UIRs of the Poincar\'e algebra}
\begin{center}
{\begin{tabular}{|c|c|c|c|}
 \hline
  UIR & \# of components & Little subgroup 
	\\\hline\hline
  \small{Discrete spin} & \small{finite} &  $SO(D-2)$ 
  \\\hline
  \small{Continuous spin} & \small{infinite} & $SO(D-3)$ 
	\\\hline
\end{tabular}
\label{Table3}
}
\end{center}
\end{table}

The subgroup $SO(D-2)\subset ISO(D-2)$  of the massless little group can be called the ``effective little group'' of helicity representations. It admits nontrivial representations for spacetime dimensions $D\geqslant 5$ only. Therefore, in four dimensions, discrete/finite-spin representations are labelled by a single real number: their helicity. 

The subgroup $SO(D-3)\subset ISO(D-2)$ of the massless little group is sometimes called the ``short little group'' of continuous-spin representations \cite{Brink:2002zx}. This group is degenerate for $D\leqslant 3$ and admits nontrivial representations for $D\geqslant 6$ only. Therefore, in four dimensions, there exist only two infinite-spin representations: the single-valued (bosonic) and the double-valued (fermionic) ones, whose physical components span all (integer or half-integer) helicities. 

\subsection{Spinning degrees of freedom}

For spacetime dimension $D=4$, the physical components forming a UIR of the little group $ISO(2)$ can be realized as square-integrable functions on the circle in the transverse plane, hence these UIRs are labelled by the radius $\mu$ of the corresponding circle. 
There are two natural bases\footnote{These two bases correspond to the position and momentum representation of these square-integrable functions on the circle.} for the spectrum of states:
\begin{itemize}
	\item ``\textbf{angle basis}'':	The states $\mid\theta\,\rangle$ are eigenstates of the null boosts (the ``translations'' in the transverse plane) but transverse rotations transform these states into each other $\mid\theta\,\rangle\to\,\mid\theta+\alpha\,\rangle$. 
	\item ``\textbf{helicity basis}'': The Fourier dual basis elements $\mid h\,\rangle$ are eigenstates of the transverse rotation generator, but the null boosts mix this infinite tower of helicity eigenstates.
\end{itemize}
Remark that the mixing of helicity eigenstates disappears in the limit $\mu\to0$ in which the continuous-spin representation becomes the direct sum of all (integer or half-integer) helicity representations. In more physical terms: at energies $E\gg\mu$, free continuous-spin particles behave as an infinite tower of particles with distinct helicities.

\subsection{Casimir operators of the Poincar\'e algebra}

Another traditional method for classifying UIRs of the Poincar\'e group is to make use of the eigenvalues of the Casimir operators. It has virtues and drawbacks: 
\begin{itemize}
	\item[+] In this approach, Lorentz covariance is more direct and the physical interpretation of Casimir operators (e.g. in $D=4$: the squares of momentum and Pauli-Lubanski vectors) may be more enlightening than the method of induced representations.
	\item[-] The UIRs of (finite-dimensional) semisimple Lie algebras are characterized uniquely by the eigenvalues of their independent Casimir operators. However, this is not necessarily true for non-semisimple Lie algebras (such as Poincar\'e algebra), there can be degeneracies (e.g. all helicity representations have vanishing quadratic and quartic Casimir operators).
\end{itemize}

The quadratic Casimir operator of the
Lorentz algebra $\mathfrak{so}(D-1,1)$ is the square of the
generators $\hat{J}_{\mu\nu}$: 
\begin{equation}
\hat{\cal C}_2\Big(\mathfrak{so}(D-1,1)\Big) \,=\,\frac12\,\hat{J}^{\mu\nu}\hat{J}_{\mu\nu}\,.
\end{equation}
The quadratic Casimir
operator of the Poincar\'e algebra $\mathfrak{iso}(D-1,1)$ is the
square of the momentum
\begin{equation}
\hat{\cal C}_2\Big(\mathfrak{iso}(D-1,1)\Big) \,=\,-\hat{P}^{\mu}\hat{P}_{\mu}\,,
\end{equation}
while the
quartic Casimir operator of the Poincar\'e algebra $\mathfrak{iso}(D-1,1)$ is
\begin{equation}
\hat{\cal C}_4\Big(\mathfrak{iso}(D-1,1)\Big) \,=\, -{1\over 2}\hat{P}^2\hat{J}_{\mu\nu}\hat{J}^{\mu\nu}
+\hat{J}_{\mu\rho}\hat{P}^\rho \hat{J}^{\mu\sigma}\hat{P}_\sigma\,,
\end{equation}
which, for $D=4$,
is the square of the Pauli-Lubanski vector, 
\begin{equation}
\hat{W}^{\mu}:=\frac{1}{2}\,\varepsilon^{\mu\nu\rho\sigma}\hat{J}_{\nu\rho}\hat{P}_{\sigma}\,.
\end{equation}

The massive and massless representations for $D=4$ spacetime dimensions are respectively characterized by the following eigenvalues:
\begin{equation}
{\cal C}_2\Big(\mathfrak{iso}(3,1)\Big)
\,=\,\left\{
      \begin{aligned}
        \,m^2
				\\
				0\,\,\,
      \end{aligned}
    \right.\,,
\qquad
{\cal C}_4\Big(\mathfrak{iso}(3,1)\Big)
\,=\,\left\{
      \begin{aligned}
        m^2\,s(s+1)
				\\
				\mu^2\quad
      \end{aligned}
    \right.\,\,.
\label{Casimirs}
\end{equation}
All helicity representations are such that both Casimir operators vanish ($m^2=0=\mu^2$). 

\begin{figure}
\begin{center}
\includegraphics[scale=0.83]{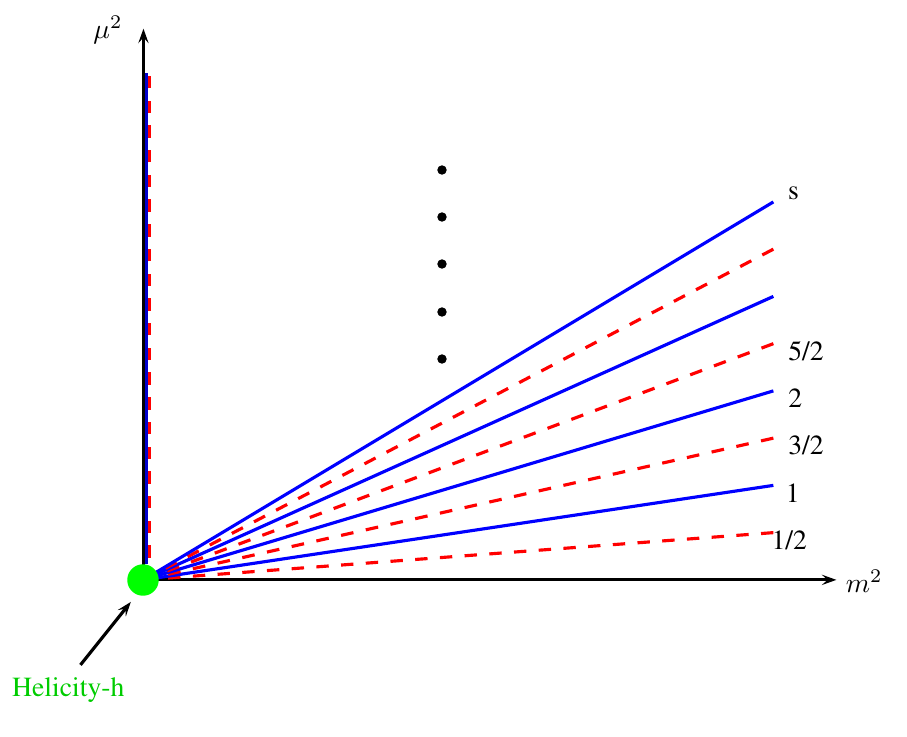}
\end{center}
\caption{Unitary irreducible representations of $\mathfrak{iso}(3,1)$ with non-negative mass squared: Single-valued UIRs sit on blue lines while double-valued ones are in dashed red. The two continuous-spin representations sit on the vertical axis. All helicity representations sit on the origin. (This figure is adapted from Fig.2 in  \cite{Schuster:2013pxj}.)
\label{Storo}}
\end{figure}

Three limits of representations are of particular interest:\footnote{It can be illuminating to visualize these limits on Fig. \ref{Storo}. The massless (respectively, helicity) limit corresponds to going towards the origin along a spin-$s$ massive line (respectively, along a vertical line). The zero-mass/infinite-spin corresponds to increasing the slope of the massive line till it becomes a vertical line.}
\begin{itemize}
	\item \textbf{Massless limit (spin fixed):} The limit $m\to 0$ of a  spin-$s$ massive representation gives the direct sum of helicity representations of spin $s$, $s-1$, $s-2$, ..., $1$, $0$\,.
	\item \textbf{Helicity limit:} The limit $\mu\to 0$ of a single continuous-spin representation (either bosonic or fermionic) gives the direct sum of an infinite tower of helicity representations (either all integer spins or all half-integer spins).
	\item \textbf{Zero-mass/Infinite-spin limit (product fixed):} The spin-$s$ massive representation becomes the continuous-spin representation (either bosonic or fermionic) in the limit \cite{Khan:2004nj}
\begin{equation}
\label{m0sinfty}
m\to0\,,\qquad s\to\infty\,,\qquad \mu=m\,s\quad\mbox{fixed}\,.
\end{equation}
\end{itemize}
The last limit is particularly interesting. It provides an interpretation of a continuous-spin particle as the high-energy ($E\gg m$) large-spin ($s\gg 1$) limit of a massive particle (in the regime $E\sim \mu=ms$) giving a simple and physical explanation of both exotic properties (i)-(ii). In particular, the energy scale $\mu$ can be seen as the remnant of the mass in the subtle massless limit \eqref{m0sinfty}.

At the level of the little group, the massless limit is the In\"{o}n\"{u}-Wigner contraction of the little groups
$SO(D-1)\,\stackrel{m\to 0}{\longrightarrow}\,\,ISO(D-2)$\,.
A geometrical interpretation of this limit is that the In\"{o}n\"{u}-Wigner contraction of groups 
$$SO(d+1)\,\stackrel{R\to \infty}{\longrightarrow}\,\,ISO(d)$$ 
corresponds to the contraction of the isometry group of the sphere in the limit of infinite radius:
$$
{\mathbb S}^{d}\,\stackrel{R\to \infty}{\longrightarrow}\,{\mathbb R}^{d}\,.
$$
At the level of representations, the simplest example of UIR of the rotation algebra $\mathfrak{so}(d+1)$ are the spherical harmonics, i.e. the solutions of the equation
\begin{equation}
\left(\Delta_{{\mathbb S}^{d}}+\frac{s(s+d-1)}{R^2}\right)Y^s_m(\vec\theta)=0
\end{equation}
whose limit
\begin{equation}
R\to\infty\,,\qquad s\to\infty\,,\qquad \mu=\,s/R\quad\mbox{fixed}
\end{equation}
is the Helmholtz equation
\begin{equation}
\left(\Delta_{{\mathbb R}^{d}}+\mu^2\right)\Phi(\vec x)=0\,,
\end{equation}
whose space of solutions carries the simplest UIR of the Euclidean algebra $\mathfrak{iso}(d)$.
This is precisely what happens at each point of spacetime in the zero-mass/infinite-spin limit: the spinning degrees of freedom of a massive particle can be realized as spherical harmonics on a sphere ${\mathbb S}^{D-2}\subset {\mathbb R}^{D-1}$ and, in the massless limit, the sphere ${\mathbb S}^{D-2}$ becomes the transverse plane ${\mathbb R}^{D-2}$.

\section{Wigner equations}\label{WEqs}

In the 1930's and 1940's, various equivalent covariant equations have been proposed (notably by Dirac, Fierz, Pauli, ...) for higher-spin fields, the solution space of which carries a massive (or helicity) UIR of the Poincar\'e group.
In 1947, Wigner proposed covariant equations, the solution space of which carries a continuous-spin UIR of the Poincar\'e group \cite{Wigner:1947}. Retrospectively, Wigner's equations can be obtained from a suitable zero-mass/infinite-spin limit of some massive higher-spin equations \cite{Bekaert:2005in}.

We will restrict the discussion to bosonic symmetric tensor fields. The reader is referred to \cite{Bekaert:2005in} for the fermionic and mixed-symmetry fields.

\subsection{Massive higher-spin equations}\label{meqs}

For integer spin $s\in{\mathbb N}$, covariant equations carrying the massive UIR of the Poincar\'e group $ISO(D-1,1)$ induced from the symmetric tensor representation of the little group $SO(D-1)$ can be formulated in terms of a symmetric Lorentz tensor:
\begin{eqnarray}
(p^2+m^2)\varphi_{\mu_1\mu_2\ldots\mu_s}(p)&=&0\label{m1}\,,\\
p^\nu\varphi_{\nu\mu_1\mu_2\ldots\mu_{s-1}}(p)&=&0\,,\\
\varphi^\nu{}_{\nu\mu_1\mu_2\ldots\mu_{s-2}}(p)&=&0\,.\label{m3}
\end{eqnarray}
Indeed, in a rest frame the second equation implies the vanishing of all timelike components while the third equation implies the $\mathfrak{so}(D-1)$-irreducibility.

A standard trick for higher-spins is to contract all indices with an auxiliary vector, say $u^\mu$, and introduce the generating function
\begin{equation}\label{spin-s}
\varphi(p,u)=\frac1{s!}\,\varphi_{\mu_1\ldots\mu_s}(p)\,u^{\mu_1}\cdots u^{\mu_s}\,
\end{equation}
so that the massive equations \eqref{m1}-\eqref{m3} read:
\begin{eqnarray}
(p^2+m^2)\varphi(p,u)&=&0\,,\label{M1}
\\
\left(p\cdot\frac{\partial}{\partial u}\right)\varphi(p,u)&=&0\,,\\
\left(\frac{\partial}{\partial u}\cdot\frac{\partial}{\partial u}\right)\varphi(p,u)&=&0\,.\label{M3}
\end{eqnarray}
To which, we should now add the homogeneity equation
\begin{equation}\label{hom}
\left(u\cdot\frac{\partial}{\partial u}\,-\,s\right)\varphi(p,u)=0\,,
\end{equation}
to keep track of the fact that the spin is fixed, i.e. equation \eqref{spin-s}. 

Unfortunately, the homogeneity equation \eqref{hom} is singular in the limit $s\to\infty$.
However, the massless limit with fixed spin is well defined, as investigated in the next subsection.

\subsection{Finite-spin/Massless limit of massive equations}

The massless limit $m\to0$ (with spin $s$ fixed) of the massive equations \eqref{M1}-\eqref{M3} are
\begin{eqnarray}
p^2\,\varphi(p,u)&=&0\,,\\
\left(p\cdot\frac{\partial}{\partial u}\right)\varphi(p,u)&=&0\,,\\
\left(\frac{\partial}{\partial u}\cdot\frac{\partial}{\partial u}\right)\varphi(p,u)&=&0\,,\\
\left(u\cdot\frac{\partial}{\partial u}\,-\,s\right)\varphi(p,u)&=&0\,.
\end{eqnarray}
However, these equations propagate too many degrees of freedom for describing a single helicity representation. 

In order for the space of solutions to carry the spin-$s$ helicity UIR of the Poincar\'e group $ISO(D-1,1)$ induced from the rank-$s$ symmetric tensor representation of the effective little group $SO(D-2)$, gauge equivalent solutions must be identified:
\begin{equation}\label{gtransf}
\varphi(p,u)\sim\varphi(p,u)+ (u\cdot p)\,\varepsilon(p,u)\,.
\end{equation}
In other words, longitudinal components are pure gauge:
\begin{equation}
\varphi_{\mu_1\ldots\mu_s}(p)\sim \varphi_{\mu_1\ldots\mu_s}(p)+p_{(\mu_1}\varepsilon_{\mu_2\ldots\mu_s)}(p)\,,
\end{equation}
where the round bracket stands for the total symmetrisation over all indices.
Strictly speaking, for consistency the gauge parameter $\varepsilon(p,u)$ should obey to similar equations
\begin{eqnarray}
p^2\,\varepsilon(p,u)&=&0\,,\label{g1}\\
\left(p\cdot\frac{\partial}{\partial u}\right)\varepsilon(p,u)&=&0\,,\\
\left(\frac{\partial}{\partial u}\cdot\frac{\partial}{\partial u}\right)\varepsilon(p,u)&=&0\,,\\
\left(\,u\cdot\frac{\partial}{\partial u}\,-\,(s-1)\,\right)\varepsilon(p,u)&=&0\,.\label{g4}
\end{eqnarray}

\subsection{Massless higher-spin equations}\label{mhseqs}

One way to get rid of the equivalence relation is to leave the space of polynomials in the auxiliary vector and reformulate the equations in terms of the gauge-invariant distribution
\begin{equation}\label{distr}
\phi(p,u)\,=\,\delta(p\cdot u)\,\varphi(p,u)\,.
\end{equation}
In terms of this distribution, the finite-spin massless equations take the form:
\begin{eqnarray}
p^2\,\phi(p,u)&=&0\label{m01}\,,\\
(p\cdot u)\,\phi(p,u)&=&0\,,\\
\left(p\cdot {\partial \over \partial u}\right) \phi(p,u)&=&0\,,\\
\left({\partial \over \partial u}\cdot{\partial \over \partial u}\right)\phi(p,u)&=&0\,,\\
\left(\,u\cdot{\partial \over \partial u}-(s-1)\,\right)\phi(p,u)&=&0\label{m04}\,.
\end{eqnarray}

Notice that this system of equations is formally identical to the one obeyed by ``reducibilities'', also called ``Killing tensor fields'' in the present case, i.e. gauge parameters $\varepsilon(p,u)$ such that the corresponding gauge transformation \eqref{gtransf} vanishes. More explicitly, the system \eqref{m01}-\eqref{m04} of five equations has the same form as the system \eqref{g1}-\eqref{g4} of four equations  supplemented by the Killing-like equation $(u\cdot p)\,\varepsilon(p,u)=0$. Naively, a paradox seems to appear: the same set of equations can describe two distinct irreducible representations of the Poincar\'e group. In the present case, the system \eqref{m01}-\eqref{m04} allows to describe either a spin-$s$ massless representation (which is infinite-dimensional and unitary) or a spin-($s-1$) Killing tensor (which span a finite-dimensional non-unitary representation). The resolution of the apparent paradox is that equations by themselves do not determine the space of their solutions, one should always specify the functional space one considers. For instance, the gauge parameter $\varepsilon(p,u)$ is polynomial in $u$ while the gauge-invariant field $\phi(p,u)$ is a distribution of the form \eqref{distr} where the gauge field $\varphi(p,u)$ is polynomial in $u$.\footnote{Strictly speaking, one should also specify the dependence in the momentum $p$. For instance, it is possible to state the functional class for Killing tensor fields in terms of their Fourier transform $\varepsilon(x,u)$, which are required to be smooth functions of the position $x$. This assumption together with the polynomiality in $p$ and the Killing equation $(\frac{\partial}{\partial x}\cdot\frac{\partial}{\partial u})\varepsilon(x,u)=0$ imply that $\varepsilon(x,u)$ is actually polynomial also in $x$.}

The effective little group of massless particles in $D+1$ dimensions is $SO(D-1)$ which coincides with the little group of massive particles in $D$ dimensions.
This is the group theoretical explanation behind the technique of dimensional reduction for obtaining massive equations from massless equations in one higher dimension, where one considers a single massive Kaluza-Klein mode.
If the higher-dimensional massless fields are gauge fields, then the dimensional reduction typically produces a tower of lower-spin fields, which are  either Stuckelberg (i.e. pure gauge) or auxiliary (i.e. which can be eliminated via their own algebraic equations of motion).
Massive equations in such a Stuckelberg approach turn out to be more convenient for taking the infinite-spin massless limit of massive equations.

\subsection{Dimensional reduction of massless equations}

Writing capital letters for the quantities in D+1 dimensions, the analogue of \eqref{m01}-\eqref{m04} reads:
\begin{eqnarray}
P^2\,\Phi(P,U)&=&0\label{M01}\,,\\
\big(P\cdot U\big)\,\Phi(P,U)&=&0\,,\\
\left(P\cdot {\partial \over \partial U}\right) \Phi(P,U)&=&0\,,\\
\left({\partial \over \partial U}\cdot{\partial \over \partial U}\right)\Phi(P,U)&=&0\,,\\
\left(\,U\cdot{\partial \over \partial U}-(s-1)\,\right)\Phi(P,U)&=&0\label{M04}\,.
\end{eqnarray}
Consider the $D+1$ splittings of the momentum and of the auxiliary vector:
\begin{equation}
P_M=(p_\mu,m)\,,\qquad U^M=(u^\mu,v)\,,
\end{equation}
the system \eqref{M01}-\eqref{M04} then reads:
\begin{eqnarray}
(p^2+m^2)\,\Phi(p,u,v)&=&0\label{Mm1}\,,\\
(p\cdot u+mv)\,\Phi(p,u,v)&=&0\,,\\
\left(p\cdot{\partial \over\partial u}+m{\partial \over \partial v}\right) \Phi(p,u,v)&=&0\,,\\
\left({\partial \over \partial u}\cdot{\partial \over \partial u}+{\partial^2 \over \partial v^2}\right)\Phi(p,u,v)&=&0\label{Mm4}\,,\\
\left(\,u\cdot{\partial \over\partial u}+v{\partial \over \partial v}-(s-1)\,\right)\Phi(p,u,v) &=&0\label{Mm5}\,,
\end{eqnarray}
and provides massive equations in D dimensions.
These massive equations are somehow a ``gauge-fixed'' version of the Stuckelberg formulation.

Let us point out that this system of equations is formally identical to the one obeyed by reducibilities in the genuine Stuckelberg formulation. For $m\neq0$, one can show (by taking derivatives and evaluating at the origin) that in the space of polynomial functions of $u$ and $v$ there are no nontrivial solutions of the system \eqref{Mm1}-\eqref{Mm5}. This is consistent with the fact that, in the initial formulation of massive higher-spin field in the section \ref{meqs}, there were no gauge symmetries. In fact, the space of reducibilities is not affected by the introduction of Stuckelberg or auxiliary fields. 

Returning to the original goal, it is clear from the equation \eqref{Mm5} that the infinite-spin limit is ill-defined in terms of the field $\Phi$.
In order to get a well defined limit, one has to extract an infinite factor
from $\Phi$ and also to assume a suitable scaling of the variable $v$, as is done in the next subsection.

\subsection{Infinite-spin/Massless limit of massive equations}

Let us introduce the parameter $\mu$ and the variable $\alpha$ by
\begin{equation}
\mu=s\,m, \quad \alpha=v/s\,.
\end{equation}
The precise limit we are interested in is: $s$ goes
to infinity, with $\mu$ and $\alpha$ both finite.

Consider the limit \eqref{m0sinfty}
and the change from the auxiliary variables $(u^\mu,v)$ to the new variables $(\omega^\mu,\alpha)$ that will be kept finite:
\begin{equation}
\left\{
      \begin{aligned}
        &u^\mu=\omega^\mu\,\alpha
				\\
				&v=s\,\alpha
      \end{aligned}
\right.
	\quad\Longleftrightarrow\quad
\left\{
      \begin{aligned}
        &\omega^\mu={s\over v}\,u^\mu
				\\
				&\alpha={v\over s}
			\end{aligned}
\right. \,\,.
\end{equation}
In fact, the problematic homogeneity condition \eqref{Mm5} can be solved as
\begin{equation}
\left(\,u\cdot{\partial \over\partial u}+v{\partial \over \partial v}-(s-1)\,\right)\,\Phi(u,v) =0
\quad\Longleftrightarrow\quad \Phi=\alpha^{s-1} \Psi\left({u\over \alpha}\right)\,,
\end{equation}
where $\Psi(\omega):=\Phi(\omega,s)$\,.
The remaining equations \eqref{Mm1}-\eqref{Mm4} can all be expressed in terms of the new field $\Psi$ which will remain finite in the limit. Before the limit, these equations still describe spin-$s$ massive fields but
their zero-mass/infinite-spin limit \eqref{m0sinfty} provides the following continuous-spin equations:
\begin{eqnarray}
p^2\,\Psi(p,\omega)&=&0\,,\label{csp1}\\
(p\cdot\omega+\mu)\,\Psi(p,\omega)&=&0\,,\label{csp2}\\
\left(p\cdot{\partial \over \partial\omega}\right)\Psi(p,\omega)&=&0\,,\\
\left({\partial \over \partial\omega}\cdot{\partial \over \partial\omega}+1\right)\Psi(p,\omega)&=&0\,.\label{csp4}
\end{eqnarray}

\subsection{Continuous-spin equations}

Performing a Fourier transform over the auxiliary vector $\omega$ leads exactly to Wigner's equations:
\begin{eqnarray}
p^2\,\widetilde{\Psi}(p,w)&=&0\label{W1}\,,\\
\big(p\cdot w\big)\,\widetilde{\Psi}(p,w)&=&0\label{W2}\,,\\
\left(p\cdot{\partial \over \partial w}-i\,\right)\widetilde{\Psi}(p,w)&=&0\label{W3}\,,\\
\big(w^2-\mu^2\big)\widetilde{\Psi}(p,w)&=&0\label{W4}\,,
\end{eqnarray}
in terms of the wave function
\begin{equation}
\widetilde\Psi(p,w)=\int d\omega\,\, \Psi(p,\omega)\,\exp\big(\,-\,i\,(w\cdot\omega)/{\mu}\big)\,.
\end{equation}
The physical components carry a UIR of the massless little group $ISO(D-2)$ because the auxiliary vector $w$ belongs to the sphere ${\mathbb S}^{D-3}\subset{\mathbb R}^{D-2}$ inside the transverse plane. 

The latter fact is not obvious. The proof goes as follows: the first equation, \eqref{W1}, obviously states that the support of the wave function is such that the momentum is lightlike. The 2nd equation, \eqref{W2}, implies that the support of the function is such that $w$ and $p$ are orthogonal. The 3rd equation, \eqref{W3}, is solved as
\begin{equation}
\widetilde{\Psi}(p\,,w+\theta\,p)\,=\,e^{i\,\theta}\,\widetilde{\Psi}(p\,,w)\,,\qquad \forall \theta\in{\mathbb R}\,,
\end{equation}
which shows that the longitudinal part of $w$ is pure gauge.
Together, the 2nd and 3rd equations, \eqref{W2}-\eqref{W3}, imply that one can assume that the auxiliary vector belongs to the transverse plane: $w\in {\mathbb R}^{D-2}$. Then the 4th equation, \eqref{W4}, leads to the conclusion we were looking for, i.e. $w\in {\mathbb S}^{D-3}$.

\section{Action principles on flat spacetime}\label{Action}

Wigner's equations, as their finite-spin massive ancestors, do not arise as Euler-Lagrange equations from an action principle.
The example of massless higher-spin fields suggests to make use of a gauge formulation. 
Indeed, gauge-invariant action principles corresponding to the helicity representations of the Poincar\'e group were written for arbitrary integer \cite{Fronsdal:1978rb} and half-integer \cite{Fang:1978wz} spin by Fang and Fronsdal. A similar formulation actually exists also for continuous-spin fields.

We will restrict our discussion to the simplest continuous-spin representation: the single-valued one with trivial representation of the short little group $SO(D-3)$.

\subsection{Fronsdal action}\label{sec:Fronsdal}
The Fronsdal equation \cite{Fronsdal:1978rb}
\begin{equation}
F_{\mu_1\cdots\mu_s}\equiv p^2\,\varphi_{\mu_1\cdots\mu_s}\,-\,p_{(\mu_1}p^\nu\varphi_{\mu_2\cdots\mu_s)\nu}\,+\,p_{(\mu_1}p_{\mu_2}\varphi_{\mu_3\cdots\mu_s)\nu}{}^\nu=0
\end{equation}
is the higher-spin generalization of Klein-Gordon (s=0), Maxwell (s=1) and linearized Ricci (s=2) equations.
It is invariant under the gauge transformations where the gauge parameter is traceless:
\begin{equation}
\delta_\varepsilon \varphi_{\mu_1\cdots\mu_s}=p_{(\mu_1}\varepsilon_{\mu_2\cdots\mu_s)}\,,\qquad \varepsilon^\nu{}_{\nu\mu_1\cdots\mu_{s-3}}=0\,.
\end{equation}
The space of double-traceless ($\varphi^{\nu\rho}{}_{\nu\rho\mu_1\cdots\mu_{s-4}}=0$)
and gauge-inequivalent solutions of Fronsdal equations carries the helicity UIR of the Poincar\'e group $ISO(D-1,1)$ induced from the symmetric tensor representation of the effective little group $SO(D-2)$.

The Fronsdal equation $F_{\mu_1\cdots\mu_s}=0$ is not variational for $s\geqslant 2$. For instance, the Ricci equation is \textit{not} the Euler-Lagrange equation of the Einstein-Hilbert action. 
But the higher-spin generalization 
\begin{equation}\label{Einstein}
G_{\mu_1\cdots\mu_s}\equiv F_{\mu_1\cdots\mu_s}-\tfrac12\, g_{(\mu_1\mu_2}F_{\mu_3\cdots\mu_s)\nu}{}^\nu=0
\end{equation}
of linearized Einstein's equation  is variational \cite{Fronsdal:1978rb}. The Lagragian is simply the contraction of the gauge field $\varphi_{\mu_1\cdots\mu_s}$ with the Einstein-like tensor $G_{\mu_1\cdots\mu_s}$.

Again it turns out to be technically convenient to make use of the generating function \eqref{spin-s}, which gives 
\begin{equation}
\left[\,p^2-(p\cdot u)\left(p\cdot
{\partial \over
\partial u}\right)+{1\over 2}\,(p\cdot u)^2 \left({\partial \over \partial
u}\cdot{\partial \over \partial u}\right)\right]\varphi(p,u)=0
\end{equation}
with the conditions
\begin{equation}
\left(u\cdot{\partial \over
\partial u}-s\right)\varphi(p,u)=0\,,\qquad\left({\partial \over
\partial u}\cdot{\partial \over \partial u}\right)^2\varphi(p,u)=0
\end{equation}
and gauge equivalence
\begin{equation}
\delta_\varepsilon \varphi(p,u)\,=\,(p\cdot u)\, \varepsilon(p,u)
\end{equation}
with
\begin{equation}
\left(u\cdot{\partial \over
\partial u}-(s-1)\,\right)\varepsilon(p,u)=0 \,,\qquad\left({\partial \over
\partial u}\cdot{\partial \over \partial u}\right)\,\varepsilon(p,u)=0\,.
\end{equation}
Performing the same steps as in the previous section (and a rescaling $\omega\to\mu^{-\frac12}\omega$), one may obtain the infinite-spin counterpart of Fronsdal's formulation via the following rules:
\begin{enumerate}
	\item remove the homogeneity conditions, 
	\item perform the following replacement $u\rightarrow \omega$, and
	\item take into account the replacement rules: 
$$p\cdot u\,\longrightarrow\, p\cdot\omega +\mu^\frac12\,,\qquad{\partial \over \partial
u}\cdot{\partial \over \partial u}\,\longrightarrow\, {\partial \over \partial
\omega}\cdot{\partial \over \partial \omega} +\mu\,.$$
\end{enumerate}
This leads to:
\begin{equation}
\left[p^2-(p\cdot\omega+\mu^\frac12)\left(p\cdot
{\partial \over
\partial \omega}\right)+{1\over 2}\,(p\cdot \omega)^2 \left({\partial \over \partial
\omega}\cdot{\partial \over \partial \omega}+\mu\right)\right]\varphi(p,\omega)=0
\end{equation}
with the conditions
\begin{equation}
\left({\partial \over\partial \omega}\cdot{\partial \over \partial \omega}+\mu\right)^2\varphi(p,\omega)=0
\end{equation}
and gauge equivalence
\begin{equation}
\delta_\varepsilon \varphi(p,\omega)\,=\,(p\cdot \omega+\mu^\frac12)\, \varepsilon(p,\omega)
\end{equation}
with
\begin{equation}
\left({\partial \over\partial \omega}\cdot{\partial \over \partial \omega}+\mu\right)\,\varepsilon(p,\omega)=0\,.
\end{equation}
These equations provide a gauge formulation of a bosonic continuous-spin field, reproducing in the limit $\mu\to0$ the Fronsdal formulation of massless higher-spin fields.

To determine the space of reducibilities in this Fronsdal-like approach is a subtle issue. Formally, the system of equations obeyed by the reducibilities takes (up to the rescaling $\omega\to\mu^{\frac12}\omega$) exactly the same form as the Wigner equations \eqref{csp1}-\eqref{csp4}.
As explained in the section \ref{mhseqs},
one should specify carefully the functional space in order to determine the precise space of reducibilities. This issue is an important one when looking for the nonlinear theory. In higher-spin gravity, the space of all Killing tensor fields on AdS, for the whole tower of gauge fields, can be endowed with a Lie algebra structure and forms the AdS higher-spin algebra, which is instrumental for introducing consistent interactions.
In the case of continuous-spin gauge fields, the issue remains elusive because there are no polynomial solutions (not even formal power series) in $\omega$ to Wigner's equations.\footnote{The proof relies on the simple fact that, for $\mu\neq0$, the operator $p\cdot \omega+\mu $ is invertible in the space of formal power series in $\omega$. Thence \eqref{csp2} has no nontrivial solution in this functional space.} In fact, the analogy between continuous-spin and massive particles of large spin suggests the absence of any reducibility. Nevertheless, the existence (or not) of a continuous-spin algebra remains an open issue in the present stage of understanding.

As far as an action principle for continuous-spin fields is concerned, unfortunately the (index-free version of) the Einstein-like equation \eqref{Einstein} blows up in the limit $s\to\infty$, even if one takes into account the infinite rescaling \cite{Bekaert:2005in}.
As explained in details in \cite{BMN} and reviewed here, one way out is to perform Fourier transforms over the auxiliary vector and solve the tracelessness constraints by distributions, thereby recovering the equations of motion presented by Schuster and Toro.

\subsection{Schuster-Toro action}

The Fourier transform over the auxiliary vector for the gauge parameter and field are:
\begin{eqnarray}
\widetilde\varepsilon(p,w)&=&\int d\omega\,\, \varepsilon(p,\omega)\,\exp(-i\,\mu^\frac12\,w\cdot\omega)\,,\\
\widetilde\varphi(p,w)&=&\int d\omega\,\, \varphi(p,\omega)\,\exp(-i\,\mu^\frac12\,w\cdot\omega)\,.
\end{eqnarray}
This Fourier transformation allows to solve the tracelessness constraints by distributions. In particular, distribution theory states that
\begin{equation}
\left( w^2 + 1 \right) \, \widetilde{\varepsilon} (w) = 0
\quad\Longleftrightarrow\quad
\widetilde{\varepsilon} (w) =  \delta(w^2 +1) \, \epsilon (w)
\end{equation}
and
\begin{equation}
\left( w^2 + 1 \right)^2 \, \widetilde{\varphi} (w) = 0
\quad\Longleftrightarrow\quad
\widetilde{\varphi} (w) =  \delta'(w^2 +1) \, \Phi (w)\,.
\end{equation}
In terms of this new field $\Phi$, the Fronsdal-like equation reads $\widehat{K} \Phi= 0$ where the kinetic operator
\begin{equation}
\widehat{K}=- \, \delta'(w^2 +1) \, p^2 + \frac{1}{2} \left( p\cdot\frac{\partial}{\partial w} -i\, \mu\right)  \delta( w^2 +1) \left(p \cdot \frac{\partial}{\partial w} -i\, \mu\right)
\end{equation}
is manifestly Hermitian, $\widehat{K}^\dagger=\widehat{K}$, with respect to the conjugation
\begin{equation}
w^\dagger= w\,,\qquad \left(\frac{\partial}{\partial w}\right)^\dagger\,=\,-\,\frac{\partial}{\partial w}\,.
\end{equation}
The gauge symmetries read
\begin{equation}
\delta_{\epsilon,\chi}\Phi\,= \left[    \,p\cdot w   -  \frac{1}{2} \left( w^2+1 \right)  \left(p \cdot\frac{\partial}{\partial w} - i \mu     \right)   \right] \epsilon
 + \frac{1}{2} \left( w^2 + 1 \right)^2 \, \chi
\end{equation}
This explains the origin of the bosonic action proposed by Schuster \& Toro \cite{Schuster:2013pta}:
\begin{eqnarray}
S[\Phi] & = & \frac{1}{2} \, { \int  d^4 x \, d^4 w }  \,\, \Phi(x, w) ~ \widehat{K}~ \Phi(x, w)\\ =&-&\frac{1}{2} \, { \int  d^4 x \, d^4 w }  \,\, \Phi \left[ {- \,\delta'( w^2 +1) \, \Box + \tfrac{1}{2} \left( \partial_{ w} \cdot \partial_x + \mu\right)  \delta( w^2 +1) \left( \partial_{ w} \cdot \partial_x +  \mu\right)}\right] \Phi\nonumber 
\end{eqnarray}
A similar action principle was proposed by Segal in 2001 for higher-spin massless fields on (anti) de Sitter spacetime \cite{Segal:2001qq}.
The above line of reasoning was applied in the fermionic case to construct the action \cite{Najafizadeh:2015uxa}, as will be presented in details in \cite{BMN}.

\section{Anti de Sitter spacetime}\label{AdS}

Tachyonic as well as continuous-spin fields seem non-unitary on de Sitter spacetime, as they are absent from the dictionary between UIRs of $SO(D,1)$ and fields on  $dS_D$ \cite{Basile:2016aen}. Similarly, they are also absent from the well-established dictionary between lowest-weight UIRs of $SO(D-1,2)$ and fields with bounded energy on anti-de Sitter space $AdS_D$ \cite{Metsaev:1995re,Metsaev:1997hi}. However, Metsaev proposed \cite{Metsaev:2016lhs,Metsaev:2017ytk} action principles, which we discuss below, for the spinning tachyonic fields as well as for continuous-spin\footnote{These fields on anti de Sitter space were referred to as ``continuous-spin fields'' in the sense that they have an infinite number of physical degrees of freedom.} fields on $(A)dS_D$. According to his analysis, continuous-spin fields can be unitary (in the sense of the absence of ghosts, i.e. negative kinetic terms in the action) on anti-de Sitter spacetime under some conditions on the two continuous parameters entering the action.

\subsection{Some representation theory}

For the sake of simplicity, let us particularize  the discussion to the case $D=4$ for a moment. The \textit{lowest-weight} UIRs of $AdS_4$ isometry algebra $\mathfrak{so}(3,2)$ = span$\{\hat{J}_{AB}\}$ ($A,B=0,1,2,3,4$) are characterized by the lowest energy $E_0$ (in units of the inverse of the curvature radius $R$) and the $\mathfrak{so}(3)$-spin $s$. The corresponding eigenvalues \cite{Evans} are for the quadratic Casimir operator
\begin{equation}
{\cal C}_2\Big(\mathfrak{so}(3,2)\Big):=\frac12\, J_{AB}J^{AB}
=
{-\tfrac94+\big(E_0+\tfrac32\big)^2}
+s(s+1)\,,
\label{C'2}
\end{equation}
and for the quartic operator (= the square of the Pauli-Lubanski vector)
\begin{equation}
{\cal C}_4\Big(\mathfrak{so}(3,2)\Big):=W_A W^A=
\Big[\,
{-\tfrac14+\big(E_0-\tfrac32\big)^2}
\,\Big]\,s(s+1)\,,
\label{C'4}
\end{equation}
where $W_A:=\frac18\,\epsilon_{ABCDE}J^{BC}J^{DE}$\,.
The In\"onu-Wigner contraction $\mathfrak{so}(3,2)\stackrel{R\to\infty}{\longrightarrow}\mathfrak{iso}(3,1)$ is implemented via the relation $P_\mu:=J_{\mu 4}/R$. At the level of the Casimir operators and eigenvalues, the relation reads
\begin{eqnarray}
&\frac1{R^2}{\cal C}_2\Big(\mathfrak{so}(3,2)\Big)\stackrel{R\to\infty}{\longrightarrow} {\cal C}_2\Big(\mathfrak{iso}(3,1)\Big)\,,&\\
&\frac1{R^2}{\cal C}_4\Big(\mathfrak{so}(3,2)\Big)\stackrel{R\to\infty}{\longrightarrow}{\cal C}_4\Big(\mathfrak{iso}(3,1)\Big)\,,&\\
&\frac{E_0}{R}\stackrel{R\to\infty}{\longrightarrow} m\,,&
\end{eqnarray}
which reproduces the eigenvalues \eqref{Casimirs} for a massive spin-$s$ representation in the flat limit.
The continuous-spin representation is obtained in the following scaling limit 
\begin{equation}\label{scaling}
E_0\sim(\mu R)^\frac12\sim s
\end{equation}
on the energy and spin. 

At finite spin and curvature, the generic situation is as follows: the eigenvalues ${\cal C}_2\big(\mathfrak{so}(3,2)\big)=(mR)^2$ and ${\cal C}_4\big(\mathfrak{so}(3,2)\big)=(\mu R)^2$ of the two Casimir operators define two energy scales $m$ and $\mu$ (on top of the curvature radius).
Consider such a massive particle on $AdS$ with the scaling behaviour \eqref{scaling}. In the limit of large spin ($s\gg 1$): 
\begin{equation}
\mu\gg m\gg \frac1{R}
\end{equation}
since $m\sim\frac{\mu}{s}$ and $\frac1{R}\sim\frac{\mu}{s^2}$.
At very large energy $E\gg\mu$, this particle cannot be distinguished from a massless particle of large spin on flat spacetime. At energies of the order $\mu$, the mixing of helicity states under Lorentz boosts becomes relevant and this particle appears as a continuous-spin particle on flat spacetime. At energies of the order $m\sim\frac{\mu}{s}$, the the particle appears as a massive particle of large spin, still on flat spacetime. It is only at energies of the order of the scalar curvature $\frac1{R}\sim\frac{\mu}{s^2}$, that the spacetime is probed as $AdS$ spacetime.

While the representation theory of continuous-spin fields in (anti)-de Sitter space has not been elaborated on, let us propose a plausible scenario, which is based on AdS/CFT inspired considerations. By the usual AdS/CFT argument, any bulk field in anti-de Sitter space (i.e. a representation of the anti-de Sitter algebra) should correspond to some conformal field on the boundary. The duality map at the free level is an intertwining operator \cite{Gunaydin:1998jc} that maps the compact slicing $\mathfrak{so}(D-1)\oplus \mathfrak{so}(2)$ of the anti-de Sitter algebra $\mathfrak{so}(D-1,2)$ to the non-compact one $\mathfrak{so}(D-2,1)\oplus \mathfrak{so}(1,1)$. Usual fields in anti-de Sitter space, i.e. massive, massless and partially-massless (see  \cite{Metsaev:1995re,Metsaev:1997nj,Deser:2001us,Boulanger:2008up,Skvortsov:2009zu} and refs therein) are lowest-weight irreducible representations. They are characterized by the weights of the maximal compact subalgebra $\mathfrak{so}(D-1) \oplus \mathfrak{so}(2) \subset \mathfrak{so}(D-1,2)$. This construction does not seem to apply to the continuous-spin case since such representations do \textit{not} appear in the known classification. A possible scenario is to induce a representation from an irreducible representation of $\mathfrak{so}(D-2,1)\oplus \mathfrak{so}(1,1)$. In its turn, representations of $\mathfrak{so}(D-2,1)$ are constructed as induced representations that are characterized by a weight of $\mathfrak{so}(D-3)$ and an additional parameter, continuous for the principal and complementary series (see e.g. \cite{Basile:2016aen} for a review of the classification). As a result, we can have two continuous parameters and the spinning degrees of freedom are specified by a representation of $\mathfrak{so}(D-3)$. This situation coincides qualitatively with the flat space counterpart.

Let us illustrate the relation between different representations of the Poincar\'e and anti-de Sitter algebra that are relevant both for continuous-spin and higher-spin theories. In the simplest case one can start with the free massless scalar field on Minkowski spacetime $\mathbb{R}^{D-2,1}$, which is a representation of the Poincar\'e algebra $\mathfrak{iso}(D-2,1)$. As is well-known, it is also a representation of the conformal algebra $\mathfrak{so}(D-1,2)$, which is the same as the  isometry algebra of anti-de Sitter spacetime $AdS_D$. This representation was dubbed ``$\mathrm{Rac}$'' by Flato and Fronsdal, who discovered the remarkable fact that \cite{Flato:1978qz} the tensor product of two $\mathrm{Rac}$'s decomposes into the sum of the anti-de Sitter algebra $\mathfrak{so}(D-1,2)$ representations corresponding to massless fields of all integer spins $s=0,1,2,...$. The flat limit, i.e. $R\rightarrow\infty$, of this spectrum is the direct sum of all massless higher-spin fields in flat spacetime, which is the same as the $\mu\rightarrow0$ limit of a continuous-spin particle:
\begin{diagram}
  \mathrm{Rac}& \rTo^{\qquad\otimes} & \mathrm{Rac}\otimes \mathrm{Rac} & \rTo^{\mathrm{Decomposition}\qquad} & \sum_s \text{HS gauge fields on AdS}\\
    &&  \rule{0pt}{14pt}
    & & \dTo_{R\rightarrow\infty}  \\
    &&\text{Continuous-Spin on flat}&      \rTo^{\mu\rightarrow0}   & \sum_s\text{HS gauge fields on flat} 
\end{diagram}
Note that the $\mathrm{Rac}$, on the upper left of the diagram, is a field living in $D-1$ dimensions while the massless higher-spin fields (both in anti-de Sitter and in flat spacetime) as well as the continuous-spin field, live in $D$ dimensions. This diagram might suggest the question: is there any analog of $\mathrm{Rac}$ for continuous-spin fields? The answer seems to be negative, since continuous-spin representations are irreducible, they correspond to elementary particles (both in flat and AdS spacetimes) which are, by definition, \textit{not} composed of a more elementary particle (the would-be flat counterpart of the ``Rac''). For this reason, it appears that one should not look for a Flato-Fronsdal theorem for continuous-spin fields, although the question remains tantalising.

\subsection{Action principles on (anti) de Sitter spacetime}\label{AdSaction}
We discuss below the progress \cite{Metsaev:2016lhs,Metsaev:2017ytk} with action principles for continuous-spin fields in anti-de Sitter space. Fronsdal fields, discussed in section \ref{sec:Fronsdal}, turn out to be a very convenient building block for massive, partially-massless  \cite{Deser:2001us} and continuous-spin fields, as we review below. 

Let us begin with the case of a massive bosonic spin-$s$ field in flat or (anti)-de Sitter space. The idea of Zinoviev \cite{Zinoviev:2001dt} is to take advantage of the fact that the helicity content of a massive spin-$s$ particle is that of a direct sum of massless particles with spins from $0$ to $s$. Of course, the representation is not a direct sum. Nevertheless, one can try to write down the most general two-derivative and quadratic in the fields action for Fronsdal tensors $\varphi_{\mu_1...\mu_k}$ ($k=0,1,...,s$) and make it gauge invariant under the most general gauge transformation law that one can construct with the help of parameters $\varepsilon_{\mu_1...\mu_n}$ ($n=0,...,s-1$). 

In practice \cite{Zinoviev:2001dt}, one takes the sum of the flat\footnote{This was the ansatz adopted in \cite{Zinoviev:2001dt}, while an equally good starting point would be to take the AdS Fronsdal actions.} space Fronsdal actions \cite{Fronsdal:1978rb} with partial derivatives $\partial$ replaced by covariant ones $\nabla$ and also one adds all possible mixing terms, which are of two types: derivative and non-derivative, i.e. mass-like terms,
\begin{align*}
    \mathcal{L}&=\sum_{k=0}^s \mathcal{L}^F(\varphi_k)+\sum_k\left[d_k\,(\varphi_k)^2+e_k\,(\varphi_k')^2+f_k\,\varphi_{k}\varphi'_{k+2}\right]+\\
    &+\sum_k(-)^k\left[a_k\, \varphi_{k-1}(\nabla\cdot \varphi_k)+b_k\, \varphi'_{k}(\nabla\cdot\varphi_{k-1})+c_k\,(\nabla\cdot \varphi_{k}')\varphi'_{k-1}\right]\,,
\end{align*}
where $a_k$, $b_k$, $c_k$, $d_k$, $e_k$, $f_k$, are real coefficients. 
Here $\mathcal{L}^F(\varphi_k)$ is the Fronsdal Lagrangian with derivatives covariantized,\footnote{The convention for the Riemann tensor is $R_{\mu\nu,\alpha\beta}=-\Lambda(g_{\mu\alpha}g_{\nu\beta}-g_{\mu\beta}g_{\nu\alpha})$} $\varphi'_k$ denotes the trace $\varphi^{\nu}{}_{\nu\mu_3...\mu_k}$, while $\nabla\cdot\,\varphi_k$ denotes the divergence $\nabla^\nu\varphi_{\nu\mu_2...\mu_k}$ and all other indices are assumed to be contracted. Also, one takes the most general form of gauge transformations
\begin{align}\label{mostgengs}
    \delta \varphi_k&= \nabla \varepsilon_{k-1}+\alpha_k\,\varepsilon_k+\beta_k\,g\,\varepsilon_{k-2}\,,
\end{align}
where $\nabla\varepsilon_{k-1}$ stands for the symmetrized covariant gradient $\nabla_{(\mu_1}\varepsilon{}_{\mu_2...\mu_k)}$ and $g\,\varepsilon_{k-2}\equiv g_{(\mu_1\mu_2}\varepsilon_{\mu_3...\mu_k)}$. Lastly, one requires the action to be gauge invariant. As a result, all the free coefficients at fixed $k$ get expressed in terms of $\alpha_k$. Furthermore, one also arrives at a single recurrence relation for the coefficients $q_k:=\alpha_k^2$\,:
\begin{align*}
     q_{k-1}\frac{2 k (D+2 k-5)}{D+2 k-6}-q_k\frac{(k+1)  (D+2 k-2)}{D+2 k-4}+(1-k) q_{k-2}-2 \Lambda  (D+2 k-5)&=0\,,
\end{align*}
where $\Lambda$ is the cosmological constant.
The most general solution of this second order recurrence relation depends on two constants. However, if ones requires that the Lagrangian truncates and does not contain fields with spin greater than $s$, then we have to impose $q_s=0$. As a result, the second constant can be related to the mass, a convenient normalization being $q_{s-1}=m^2/s$ (as it will be clear below, $m^2=0$ corresponds to the massless spin-$s$ field). The solution is \cite{Zinoviev:2001dt}
\begin{align}\label{solzin}
    q_k&=\frac{(s-k) (D+k+s-3) \left(m^2-\Lambda  (s-k-1) (D+k+s-4)\right)}{(k+1) (D+2 k-2)}
\end{align}
for $k=0,1,...,s$\,.
The gauge invariance of the action facilitates the counting of physical degrees of freedom: they indeed correspond to a rank-$s$ irreducible representation of $\mathfrak{so}(d-1)$. Another way to see that the action describes a massive spin-$s$ field is to use the algebraic (shift) part of the gauge transformations \eqref{mostgengs}, to be precise the second term there, in order to set to zero $\varphi_0$, $\varphi_1$ and the traceless part of the $\varphi_k$ ($k=2,...,s-1$) with the help of the parameters $\varepsilon_k$ ($k=0,...,s-1$). As a result, one recovers (in flat space) the field content of the Singh-Hagen action \cite{Singh:1974qz} for a single massive spin-$s$ field.

The zoo of irreducible fields in (anti)-de Sitter space is more diverse as compared to the flat space and understanding various limits is important. In particular, there exist partially-massless fields \cite{Deser:2001us}. In the simplest case of totally-symmetric fields, a partially-massless field is specified by its spin $s$ and the depth $t$ of partial masslessness, the gauge transformation law having higher derivatives
\begin{align}
    \delta \varphi_{\mu_1...\mu_s}&=\nabla_{(\mu_1}\cdots\,\nabla_{\mu_{t}}\varepsilon_{\mu_{t+1}...\mu_s)}\,.
\end{align}
For $t=1$ we have the usual massless fields, but all integers $t$ in the range $1,...,s$ are allowed. The number of physical degrees of freedom of partially-massless fields lies in between those of massive and massless ones. In particular, in the case of spacetime dimension $D=4$ a spin-$s$ and depth-$t$ partially-massless field possesses helicities in the range $\pm(s-t+1),...\,,\pm\, s$\,. 

The Zinoviev action also includes all partially-massless cases. Indeed, by fine-tuning $m^2$ in \eqref{solzin} we can get $q_{s-t}=0$ for 
\begin{align}
    m^2_{s,t}=\Lambda (t-1)  (D+2s-t-4)\,, && (t=1,...,s)
\end{align}
and the action splits into two parts: the fields $\varphi_{s-t+1},...,\varphi_{s}$ describe a partially-massless field of spin-$s$ and depth-$t$ while the fields $\varphi_0,...,\varphi_{s-t}$ describe a non-unitary massive field of spin $s-t$. There is also a truncation to the massless field for $t=1$, where one is left with the Fronsdal action with an appropriate mass-like term. Frame-like Lagrangians for partially-massless \cite{Skvortsov:2006at} and massive \cite{Ponomarev:2010st} fields were also constructed following the procedure of Zinoviev.

The idea of the Metsaev action \cite{Metsaev:2016lhs} for continuous-spin fields can be interpreted as follows: we do not have to fix the free parameters so that $q_s=0$. Rather, one can keep the two free constants as genuine parameters of a solution. As a result, the most general solution is
\begin{align}\label{solcspA}
\begin{aligned}
   q_k&= \frac{(s'-k) (s'+D+k-3)} {(k+1) (s'-s+1) (D+2 k-2) (s'+D+s-4)}\times\\
   &\quad\times\left(\Lambda  (s'-s+1) (k-s+1) (s'+D+s-4) (D+k+s-4)+m^2 (D+2 s-4)\right)\,.
\end{aligned}   
\end{align}
It reduces to \eqref{solzin} if we set $s'=s$. Here we expressed the first constant as a condition that $q_{s'}=0$ where $s'$ does not have to be an integer. Therefore, in this technical sense, one can think of continuous-spin fields as of particles with fractional spin (see also \cite{Schuster:2014xja}\,). This analogy is not to be taken literally, continuous-spin have standard statistics (either bosonic or fermionic).

An even simpler form is obtained if we associate the two constants with two points where the function $q_k$ goes to zero, $q_{s_1}=q_{s_2}=0$: 
\begin{align}\label{solcspB}
   q_k&= - \frac{\Lambda  \left(k-s_1\right) \left(k-s_2\right) \left(D+k+s_1-3\right) \left(D+k+s_2-3\right)}{(k+1) (D+2 k-2)}\,.
\end{align}
We again would like to stress that $s_1$, $s_2$ are two formal parameters for which the two constants were traded for and they do not have to be integers. As a consequence, the generic gauge invariant action does not stop at $\varphi_s$ and contains instead Fronsdal fields with all spins $s=0,1,2,...$ This action depends on two parameters, such as $s'$ and $m^2$ (or, equivalently, $s_1$ and $s_2$). Whenever $s_1$ and/or $s_2$ is an integer we have a clear interpretation: the action splits into several parts. Solution \eqref{solcspB} is suitable for $AdS$ space, while in the flat space one should introduce at least one dimensionful parameter instead of $\Lambda$.

The crucial novelty of the most general solution, obtained by Metsaev, i.e. equivalently \eqref{solcspA} or \eqref{solcspB}, is the presence of the two free parameters, which should correspond to the two Casimir operators of the previous subsection. At generic point in the parameter space one has an irreducible system that depends on all symmetric tensor fields $\varphi_{k}$ ($k=0,1,2,...$). It is difficult to determine at present which solutions correspond to unitary representations. Nevertheless, one can use a simple reasoning as to single out the cases that are definitely \textit{not} unitary. Indeed, note that $q_k=\alpha_k^2$ and therefore unitarity requires $q_k\geqslant0$. This way it is easy to see that  partially-massless fields, i.e. $t>1$, are non-unitary in anti-de Sitter but may be (and actually are) unitary in de Sitter.

It is also clear that the Lagrangian splits into two (or three) parts whenever $s_{1}$ or $s_2$ is (or, respectively, both of them are)\footnote{The roots are exchanged under $s_i\rightarrow -D-3-s_i$. } non-negative integer(s), which allows one to see various massive, partially-massless and massless truncations. For example, the massless case corresponds to $q_{s}=q_{s-1}=0$ and the action splits into three parts for fields with spins $0,...,s-1$,  the massless field with spin $s$ and the rest with spins $s+1,s+2,...$.

\begin{figure}
\begin{center}
\parbox{4.1cm}{\includegraphics[scale=0.23]{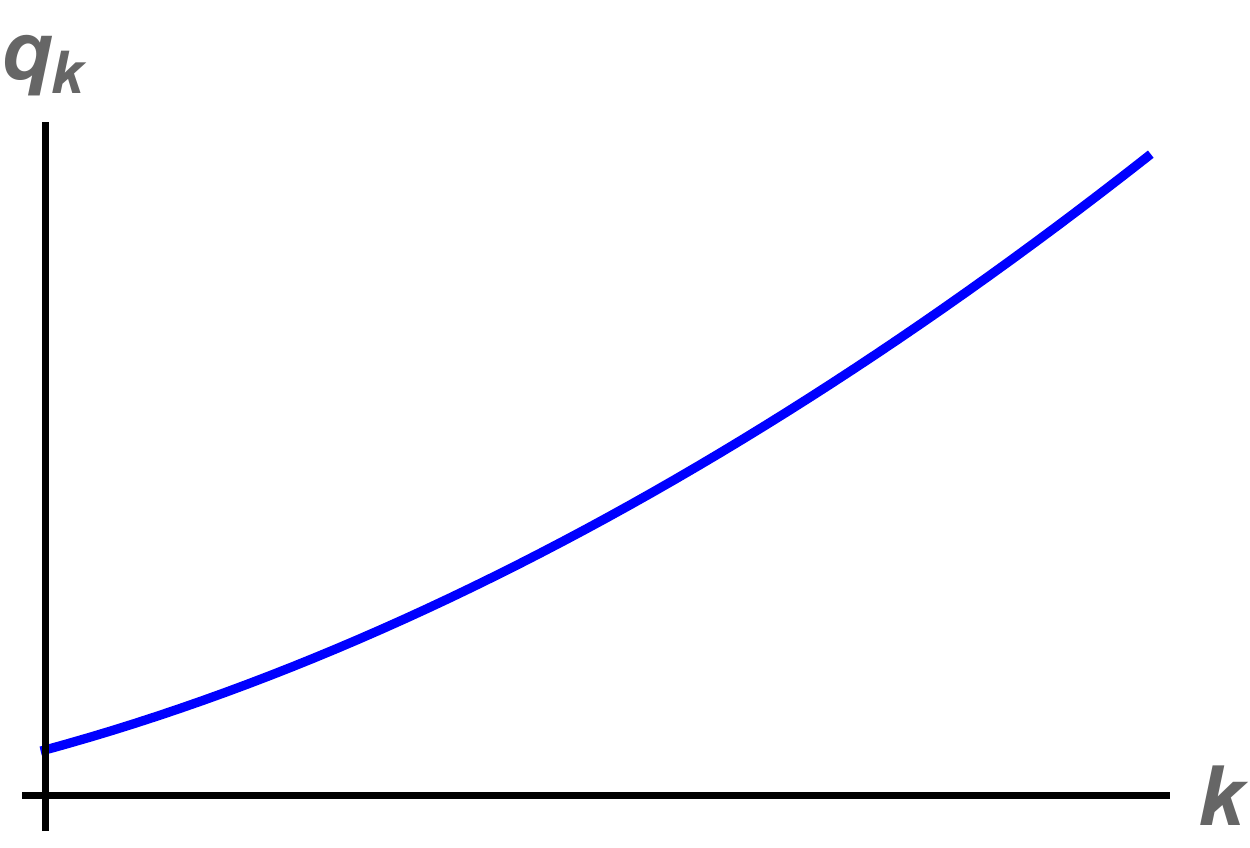}}
\parbox{4.1cm}{\includegraphics[scale=0.23]{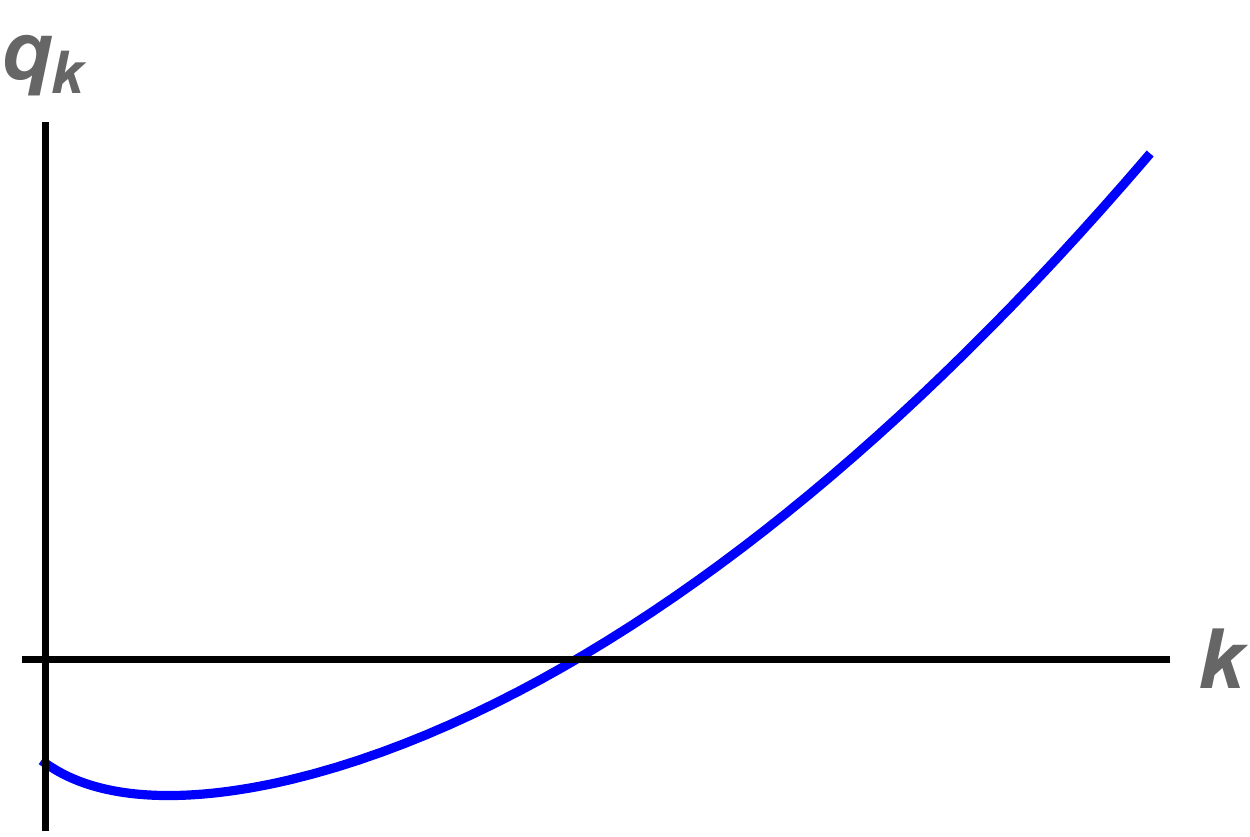}}
\parbox{4.1cm}{\includegraphics[scale=0.23]{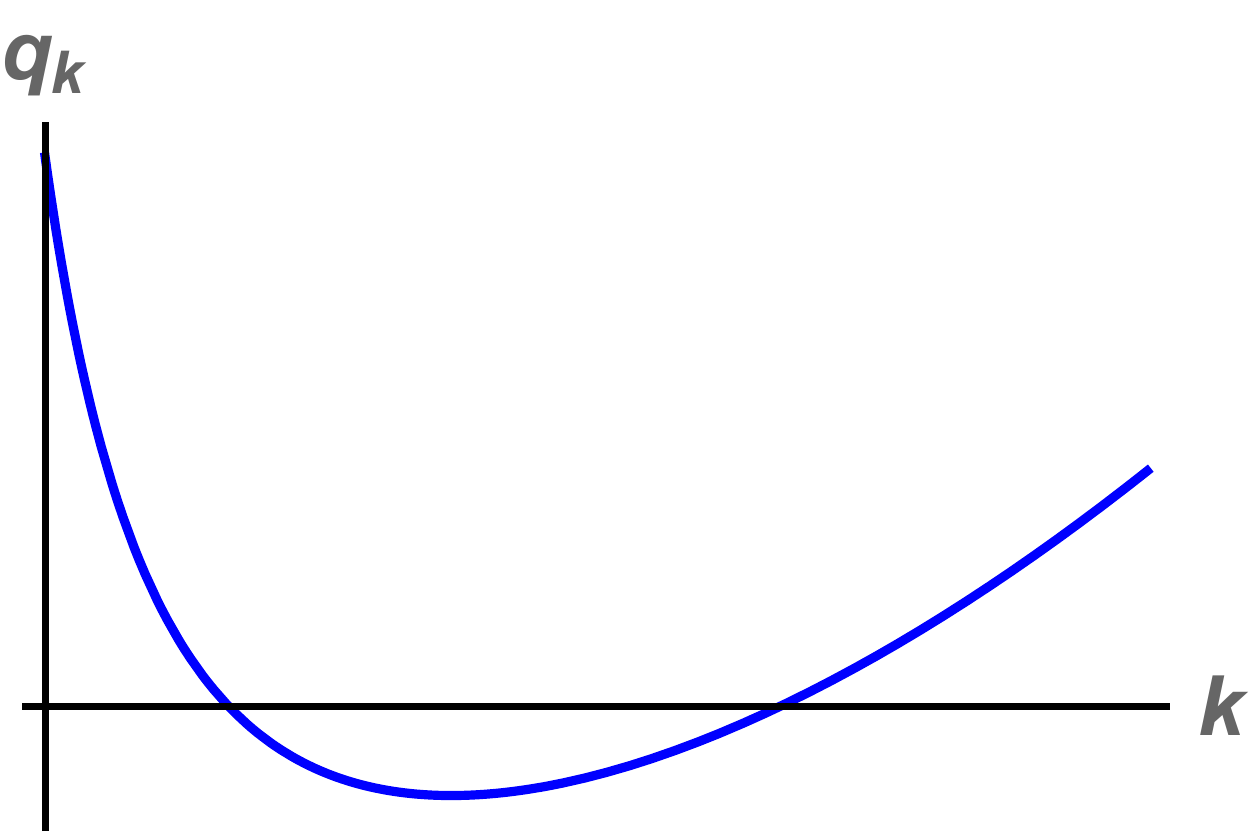}}
\end{center}
\label{fig:plots}
\caption{Typical plots of $q_k$ as a function of $k$ for various values of $s_1$ and $s_2$. The regions where $q_k$ is negative correspond to violation of unitarity unless the root is an integer and then the action splits into two (or three) parts. Those parts that have $q_k>0$ can be unitary. }
\end{figure}

The analysis of \eqref{solcspB} is very simple: there are explicitly four roots and, depending on the position of the roots, $q_k$ is either positive for all $k\geqslant0$, or can become negative for some $k$. For the de Sitter case $\Lambda>0$, we immediately see that \eqref{solcspB} goes negative for sufficiently large $k$ and therefore, continuous-spin fields cannot be unitary in de Sitter (as confirmed by their absence in the known classification of UIRs).  In the anti-de Sitter case $q_k$ is always positive for sufficiently large $k$. Therefore, if there is a range where $q_k$ goes negative for $k\geqslant0$ then we should arrange one (or two) of the roots to be integers, so that the action can split into two (or three) parts, with unitarity preserved by at least one of them. When the action splits into two parts, the first one for the finite number of fields $s=0,1,...,s'$ describes a massive spin-$s'$ field. Whenever the action splits into three parts the middle part that contains fields with spins $s'-t+1,...,s'$ describes a spin-$s'$ depth-$t$ partially-massless field. On Fig. \ref{fig:plots} we showed several plots of $q_k$ for different values of $s_1$ and $s_2$.

\subsection{Some remarks}

There is one crucial difference between continuous-spin fields in flat and anti-de Sitter spaces. In flat spacetime, there is a limit where a continuous-spin particle reduces to a direct sum of massless particles of all spins, which can be seen either from representation theory (see section \ref{Wclass}) or from the action (that splits into a direct sum of Fronsdal actions). The latter fact suggests that constructing interactions for continuous-spin fields should be closely related to higher-spin gauge theories in flat space, if any  \cite{Metsaev:unpublished,Metsaev:1991nb,Metsaev:1991mt,Taronna:2011kt,Ponomarev:2016lrm,Roiban:2017iqg,Sleight:2016xqq,Taronna:2017wbx}. On the contrary, in anti-de Sitter spacetime the maximal decomposition of the continuous-spin action we can achieve is splitting into three parts and there is no limit where one finds a direct sum of Fronsdal actions. At group-theoretical level, the UIRs of $\mathfrak{so}(D-1,2)$ corresponding to massless spin-$s$ fields saturate the unitarity bound: they are labeled by the energy $E_0=s+D-3$. The point is that their Casimir operators are distinct\footnote{For instance, \eqref{C'2}-\eqref{C'4} for the case $D=4$ with $E_0=s+1$ give ${\cal  C}_2\big(\mathfrak{so}(3,2)\big)=2(s+2)(s+1)$ and ${\cal  C}_4\big(\mathfrak{so}(3,2)\big)=s^2(s+1)(s-1)$.} for distinct $s$, in contrast with the flat case where they vanish for \textit{all} helicity representations. Therefore, continuous-spin fields on AdS do not seem to be related  in any obvious sense to AdS higher-spin gravity theories. 
In other words, in the diagram below there is no upper arrow relating the left and right upper corners:
\begin{diagram}
   \text{CS field on AdS}\quad & 
   & \quad\sum \text{HS gauge fields on AdS}\\
   \dTo_{\Lambda \rightarrow0} & & \dTo_{\Lambda \rightarrow0}  \\
    \text{CS field on flat}\quad&      \rTo^{\mu\rightarrow0}   & \quad\sum\text{HS gauge fields on flat} 
\end{diagram}

Let us also note that the 
unfolded formulation of a massive spin-$s$ field was also constructed \cite{Ponomarev:2010st}. This was done by following Zinoviev's idea that a massive spin-$s$ field can be described as an appropriate mixture of massless fields with spins $0,1,...,s$\,. Likewise, in order to construct the frame-like action of a massive spin-$s$ field one should take the field content of unfolded formulations of massless fields with spins $0,1,...,s$ and add all possible mixing terms. In practice, as in the case of the Zinoviev action, one has to solve certain recurrence equations and the most general solution was found \cite{Ponomarev:2010st}. It depends on two parameters, which is in accordance with the discussion above. If one of these parameters is taken to be an integer, then the solution truncates at some spin and the second parameter can be associated with the mass.

Another extension along the same lines is to explore continuous-spin fields of mixed-symmetry type.\footnote{Mixed-symmetry continuous-spin fields have been discussed at the group-theoretical level in  \cite{Brink:2002zx} and at the level of wave equations in  \cite{Bekaert:2005in}.} The idea is the same: continuous-spin particles can be thought of as an infinite-spin limit of massive ones. In particular, one takes the action \cite{Zinoviev:2009gh} for massive field whose spin is described by a two-row Young diagram $(s_1,s_2)$, $s_1\geqslant s_2$, and formally let $s_1$ to be non-integer \cite{Zinoviev:2017rnj}. An obstacle that prevents one from studying the most general case of mixed-symmetry continuous-spin fields is the lack of a Lagrangian formulation for massive fields whose spin is characterized by an arbitrary Young diagram, i.e. a representation of $\mathfrak{so}(D-1)$. 

\subsection{Vacuum Partition Function}\label{sec:part}
Vacuum partition function of continuous-spin fields have also been studied  \cite{Metsaev:2016lhs,Metsaev:2017ytk}. The results are similar to those for massless higher-spin multiplets in flat \cite{Beccaria:2015vaa} and anti-de Sitter space \cite{Giombi:2013fka,Giombi:2014iua,Skvortsov:2017ldz}.

In flat space there is a heuristic arg    ument for why the vacuum partition function of higher-spin theories has to be one \cite{Beccaria:2015vaa}. For free fields we easily find
\begin{align}\label{partA}
    Z_{\text{1-loop}}&=\frac{1}{\det^{1/2}_{0}|-\partial^2|}\prod_{s>0} \frac{\det^{1/2}_{s-1,\perp}|-\partial^2 |}{\det^{1/2}_{s,\perp}|-\partial^2|}\,,
\end{align}
where $\det_{\perp}$ is the determinant on the space of transverse and traceless tensors. This representation is obtained from a partition function of a spin-$s$ massless field
\begin{align}\label{partB}
    Z_{s}&=\frac{\det^{1/2}_{s-1}|-\partial^2 |\det^{1/2}_{s-1}|-\partial^2 |}{\det^{1/2}_{s}|-\partial^2| \det^{1/2}_{s-2}|-\partial^2|}\,,
\end{align}
where $\det$ is the determinant on the space of traceless tensors. The partition function corresponds to the decomposition of the Fronsdal field into traceless tensors of ranks $s$ and $s-2$, while the enumerator accounts for the gauge symmetries. In order to get \eqref{partA} from \eqref{partB} one needs to use $\det_{k}=\det_{k,\perp}\det_{k-1}$.

Assuming that we can cancel the neighbours in the infinite product \eqref{partA} we observe an exact cancellation, thence $Z_{\text{1-loop}}=1$. Similar arguments can be applied to a continuous-spin field in flat spacetime.\footnote{These cancellations are not unrelated to the formal similarity (observed in the section \ref{WEqs}) between the equations obeyed by the wave function and by the gauge parameter on shell.} The situation in anti-de Sitter spacetime is more complicated. 

At the free level the partition function is given by the product of determinants of the Fronsdal kinetic terms divided by the determinants of the ghosts associated with gauge symmetries. As a result one gets \cite{Giombi:2013fka,Giombi:2014iua}
\begin{align}
    Z_{\text{1-loop}}&=\frac{1}{\det^{1/2}_{0}|-\nabla^2-m^2_0|}\prod_{s>0} \frac{\det^{1/2}_{s-1,\perp}|-\nabla^2+M^2_{s} |}{\det^{1/2}_{s,\perp}|-\nabla^2 +m_s^2|}\,,\\
    &\quad m^2_s=\Lambda[(s-2)(D+s-3)-s]\,,\\
    &\quad M^2_{s}=\Lambda(s-1)(D+s-3)\,,
\end{align}
where $m^2_s$ is the mass-like term of a massless spin-$s$ field and $M^2_s$ is the one for its ghost. Differently from the flat case, trivial cancellation of the neighbours in the product is impossible since the mass-like term of the ghost does not coincide with that of a spin-$(s-1)$ field: $M^2_s\neq m^2_{s-1}$. Nevertheless, it can be shown \cite{Giombi:2013fka,Giombi:2014iua,Skvortsov:2017ldz} that $Z_{\text{1-loop}}=1$ for the spectrum containing all integer spins $s=0,1,2,...$.

Continuous-spin fields on AdS can also be shown \cite{Metsaev:2016lhs,Metsaev:2017ytk} to have $Z_{\text{1-loop}}=1$. This is similar to the flat space counterpart:
\begin{align}
    Z_{\text{1-loop}}&=\prod_{k=0}^{\infty} \frac{\det^{1/2}_{k-1,\perp}|-\nabla^2+\mathcal{M}^2_{k-1} |}{\det^{1/2}_{k,\perp}|-\nabla^2 +\mathcal{M}^2_k|}\,,\\
    &\quad \mathcal{M}^2_k=-\mu_0-\Lambda[k(k+D-1)+2D-4]\,,
\end{align}
where, remarkably, the same $\mathcal{M}^2_k$ corresponds to the mass-like terms for the fields and ghosts. Here $\mu_0$ is one of the free dimensionful parameters in the solution. Upon cancelling the neighbours in the product one finds $Z_{\text{1-loop}}=1$. The result for a single massless spin-$s$ field, which is a consistent truncation, can be obtained by noticing that this truncation has $\mu_0=-2\Lambda s(s+d-3)$. Then, $\mathcal{M}^2_s=m^2_s$ and $\mathcal{M}^2_{s-1}=M^2_{s-1}$.

Note that the simplest higher-spin multiplets contain either all even spins (``minimal Type A'' theory) or all integer spins (so-called ``non-minimal Type A''). There does not seem to be an analog of the minimal Type-A multiplet for continuous-spin particles since they must contain all integer helicities (consistently with their interpretation as the infinite-spin limit of a massive particle).

\section{List of open problems}\label{conclu}

To conclude, let us present a list of some open problems:
\begin{enumerate}
	\item Action principles:
		\begin{itemize}
		  \item Determine the precise relation (if any) between the Fronsdal-like actions of Metsaev  \cite{Metsaev:2016lhs,Metsaev:2017ytk} and the Segal-like actions \cite{Schuster:2013pta,Najafizadeh:2015uxa};\footnote{While this work was completed, paper \cite{MN} appeared that addresses this issue.}
			\item Generalize those actions to mixed-symmetry representations (relevant for spacetime dimensions $D\geqslant 7$), where another problem still open is how to construct the action for the most general massive mixed-symmetry field.
		\end{itemize}
	\item Consistency of scattering:
	\begin{itemize}
		\item Clarify the current exchanges obtained from known action principles;\footnote{Some detailed investigation reveal that the issue is a subtle one \cite{BMN}.}
		\item Extend the Weinberg soft theorems as to arrive at definite (either no-go or yes-go) conclusions.
	\end{itemize}
	\item Classification of consistent self-interactions in flat spacetime:
	\begin{itemize}
		\item Cubic order: analyze whether there exist consistent vertices\footnote{We thank Ruslan Metsaev for communicating us that according to his analysis, consistent cubic vertices appear to exist, which is a very encouraging result \cite{unpubl}. See, however, \cite{Schroer:2015rct}.}; 
		\item Quartic and higher orders: extend this analysis to quartic order and, hopefully, to all orders. 
		
		Here the easiest way seems to apply the light-cone approach \cite{Metsaev:1991mt} along the lines of the usual massless fields.\footnote{One can expect a continuous-spin analogue of the chiral higher-spin theory \cite{Metsaev:unpublished,Metsaev:1991nb,Metsaev:1991mt,Ponomarev:2016lrm} to exist.} However, in the covariant approach the crucial question is what is the underlying gauge algebra, if any.
	\end{itemize}
	\item Nonvanishing cosmological constant:	
	\begin{itemize}
		\item It is still important to understand the group-theoretical origin of continuous-spin UIRs of anti de Sitter isometry algebra $\mathfrak{so}(D-1,2)$;
		\item A related issue is the holographic interpretation of bulk continuous-spin gauge fields as boundary conformal operators; 
        \item If a continuous-spin algebra exists (this question was raised above), then it would be natural to ask whether it arises from an exotic nontrivial flat limit of AdS higher-spin algebra.
	\end{itemize}
\end{enumerate}

\section*{Acknowledgments}

X.B. thanks A. Campoleoni, D. Francia, E. Joung, M. Taronna for useful discussions and, in particular, J. Mourad and M. Najafizadeh for collaboration on continuous-spin fields. E.S. acknowledges D. Ponomarev, R. Metsaev and Y. Zinoviev for useful discussions. X.B. and E.S. would like to thank K.-H. Rehren and B. Schroer for correspondence. X.B. is also grateful, for hospitality, to the Lebedev Physical Institute where part of this work has been presented as colloquium and lectures.

The work of X.B. and E.S. was supported in part by the Russian Science Foundation grant 14-42-00047 
in association with Lebedev Physical Institute. The work of E.S. was supported by the DFG Transregional Collaborative 
Research Centre TRR 33 and the DFG cluster of excellence ``Origin and Structure of the Universe". 



\setstretch{0.95}


\begin{thebibliography}{0}    

\bibitem{Wigner:1939cj} 
  E.~P.~Wigner,
  ``On Unitary Representations of the Inhomogeneous Lorentz Group,''
  Annals Math.\  {\bf 40}, 149 (1939)
  [Nucl.\ Phys.\ Proc.\ Suppl.\  {\bf 6}, 9 (1989)].
	
\bibitem{Bargmann:1948ck} 
  V.~Bargmann and E.~P.~Wigner,
  ``Group Theoretical Discussion of Relativistic Wave Equations,''
  Proc.\ Nat.\ Acad.\ Sci.\  {\bf 34}, 211 (1948).
	
\bibitem{Wigner:1963in}
	E.~P. Wigner, ``Invariant quantum mechanical equations of motion,'' in \textit{Theoretical Physics Lectures presented in Trieste} (International {A}tomic {E}nergy {A}gency, Vienna, 1963) 59.
	
\bibitem{Bekaert:2005in} 
  X.~Bekaert and J.~Mourad,
  ``The continuous-spin limit of higher-spin field equations,''
  JHEP {\bf 0601}, 115 (2006)  [hep-th/0509092].
	
\bibitem{Bekaert:2010hw} 
  X.~Bekaert, N.~Boulanger and P.~Sundell,
  ``How higher-spin gravity surpasses the spin two barrier: no-go theorems versus yes-go examples,''
  Rev.\ Mod.\ Phys.\  {\bf 84}, 987 (2012)
  [arXiv:1007.0435 [hep-th]].

\bibitem{Rahman:2013sta} 
  R.~Rahman,
  ``Higher Spin Theory - Part I,''
  PoS ModaveVIII {\bf 004} (2012)
  [arXiv:1307.3199 [hep-th]].
	
\bibitem{Fradkin:1986qy}
  E.~S.~Fradkin and M.~A.~Vasiliev,
  ``Cubic Interaction in Extended Theories of Massless Higher Spin Fields,''
  Nucl.\ Phys.\ B {\bf 291}, 141 (1987).

\bibitem{Schuster:2013pxj} 
  P.~Schuster and N.~Toro,
  ``On the Theory of Continuous-Spin Particles: Wavefunctions and Soft-Factor Scattering Amplitudes,''
  JHEP {\bf 1309}, 104 (2013)
  [arXiv:1302.1198 [hep-th]].
	
\bibitem{Schuster:2013vpr} 
  P.~Schuster and N.~Toro,
  ``On the Theory of Continuous-Spin Particles: Helicity Correspondence in Radiation and Forces,''
  JHEP {\bf 1309}, 105 (2013)
  [arXiv:1302.1577 [hep-th]].

\bibitem{Weinberg:1964ew} 
  S.~Weinberg,
  ``Photons and Gravitons in s Matrix Theory: Derivation of Charge Conservation and Equality of Gravitational and Inertial Mass,''
  Phys.\ Rev.\  {\bf 135}, B1049 (1964).

\bibitem{Mund:2004sy} 
 R.~Brunetti, D.~Guido and R.~Longo,
 ``Modular localization and Wigner particles,''
 Rev. Math. Phys. {bf 14}, 759 (2002)
 [math-ph/0203021].
 J.~Mund, B.~Schroer and J.~Yngvason,
  ``String localized quantum fields from Wigner representations,''
  Phys.\ Lett.\ B {\bf 596}, 156 (2004)
  [math-ph/0402043];
  R.~Longo, V.~Morinelli and K.~H.~Rehren,
 ``Where Infinite Spin Particles Are Localizable,''
 Commun. Math. Phys. {bf 345}, no. 2, 587 (2016)
 [arXiv:1505.01759 [math-ph]].
 
\bibitem{Schroer:2015rct} 
  B.~Schroer,
  ``Wigner’s infinite spin representations and inert matter,''
  Eur.\ Phys.\ J.\ C {\bf 77}, no. 6, 362 (2017)
  [arXiv:1601.02477 [physics.gen-ph]].
   
 

\bibitem{Wigner:1947}
	E.~P.~Wigner, Z.\ Physik {\bf 124},  665 (1947).
	
\bibitem{Fronsdal:1978rb} 
  C.~Fronsdal,
  ``Massless Fields with Integer Spin,''
  Phys.\ Rev.\ D {\bf 18}, 3624 (1978).
	
\bibitem{Fang:1978wz} 
  J.~Fang and C.~Fronsdal,
  ``Massless Fields with Half Integral Spin,''
  Phys.\ Rev.\ D {\bf 18}, 3630 (1978).
	
\bibitem{Schuster:2013pta} 
  P.~Schuster and N.~Toro,
  ``A Gauge Field Theory of Continuous-Spin Particles,''
  JHEP {\bf 1310}, 061 (2013)
  [arXiv:1302.3225 [hep-th]].
	
\bibitem{Najafizadeh:2015uxa} 
  X.~Bekaert, M.~Najafizadeh and M.~R.~Setare,
  ``A gauge field theory of fermionic Continuous-Spin Particles,''
  Phys.\ Lett.\ B {\bf 760}, 320 (2016)
  [arXiv:1506.00973 [hep-th]].
	
\bibitem{Segal:2001qq} 
  A.~Y.~Segal,
  ``A Generating formulation for free higher spin massless fields,''
  hep-th/0103028.

\bibitem{Rivelles:2014fsa} 
  V.~O.~Rivelles,
  ``Gauge Theory Formulations for Continuous and Higher Spin Fields,''
  Phys.\ Rev.\ D {\bf 91}, no. 12, 125035 (2015)
  [arXiv:1408.3576 [hep-th]] \&
  ``Remarks on a Gauge Theory for continuous-spin Particles,''
  Eur.\ Phys.\ J.\ C {\bf 77} (2017) no.7,  433
  [arXiv:1607.01316 [hep-th]].
	
\bibitem{Metsaev:2016lhs} 
  R.~R.~Metsaev,
  ``continuous-spin gauge field in (A)dS space,''
  Phys.\ Lett.\ B {\bf 767}, 458 (2017)
  [arXiv:1610.00657 [hep-th]].
  
\bibitem{Metsaev:2017ytk} 
  R.~R.~Metsaev,
  ``Fermionic continuous-spin gauge field in (A)dS space,''
  arXiv:1703.05780 [hep-th].

\bibitem{Fronsdal:1978vb} 
  C.~Fronsdal,
  ``Singletons and Massless, Integral Spin Fields on de Sitter Space (Elementary Particles in a Curved Space. 7.,''
  Phys.\ Rev.\ D {\bf 20}, 848 (1979).
	
\bibitem{Fang:1979hq} 
  J.~Fang and C.~Fronsdal,
  ``Massless, Half Integer Spin Fields in De Sitter Space,''
  Phys.\ Rev.\ D {\bf 22}, 1361 (1980).
	
\bibitem{Zinoviev:2001dt} 
  Y.~M.~Zinoviev,
  ``On massive high spin particles in AdS,''
  hep-th/0108192.

\bibitem{Skvortsov:2006at} 
  E.~D.~Skvortsov and M.~A.~Vasiliev,
  ``Geometric formulation for partially massless fields,''
  Nucl.\ Phys.\ B {\bf 756}, 117 (2006)
  [hep-th/0601095].
  
\bibitem{Ponomarev:2010st} 
  D.~S.~Ponomarev and M.~A.~Vasiliev,
  ``Frame-Like Action and Unfolded Formulation for Massive Higher-Spin Fields,''
  Nucl.\ Phys.\ B {\bf 839}, 466 (2010)
  [arXiv:1001.0062 [hep-th]].
  
\bibitem{BRSTpack} 
  J.~Mourad,
  ``Continuous spin particles from a tensionless string theory,''
  AIP Conf.\ Proc.\  {\bf 861}, 436 (2006) [based on hep-th/0410009 and hep-th/0504118];
  L.~Edgren, R.~Marnelius and P.~Salomonson,
  ``Infinite spin particles,''
  JHEP {\bf 0505}, 002 (2005)
  [hep-th/0503136];
  L.~Edgren and R.~Marnelius,
  ``Covariant quantization of infinite spin particle models, and higher order gauge theories,''
  JHEP {\bf 0605}, 018 (2006)
  [hep-th/0602088];
   A.~K.~H.~Bengtsson,
  ``BRST Theory for Continuous Spin,''
  JHEP {\bf 1310}, 108 (2013)
  [arXiv:1303.3799 [hep-th]].
  
\bibitem{Francia:2007qt} 
  D.~Francia, J.~Mourad and A.~Sagnotti,
  ``Current Exchanges and Unconstrained Higher Spins,''
  Nucl.\ Phys.\ B {\bf 773}, 203 (2007)
  [hep-th/0701163].
  
\bibitem{Bekaert:2006py} 
  X.~Bekaert and N.~Boulanger,
  ``The Unitary representations of the Poincare group in any spacetime dimension,'' in the proceedings of the \textit{2nd Modave Summer School in Theoretical Physics} (6-12 Aug 2006, Modave, Belgium)
  [hep-th/0611263].
	
\bibitem{Brink:2002zx} 
  L.~Brink, A.~M.~Khan, P.~Ramond and X.~z.~Xiong,
  ``Continuous-spin representations of the Poincare and superPoincare groups,''
  J.\ Math.\ Phys.\  {\bf 43}, 6279 (2002)
  [hep-th/0205145].

\bibitem{Khan:2004nj} 
  A.~M.~Khan and P.~Ramond,
  ``Continuous-spin representations from group contraction,''
  J.\ Math.\ Phys.\  {\bf 46}, 053515 (2005);
  Erratum: [J.\ Math.\ Phys.\  {\bf 46}, 079901 (2005)]
  [hep-th/0410107].
  
\bibitem{BMN}
  X.~Bekaert, J.~Mourad, M. Najafizadeh, to appear.
	
\bibitem{Basile:2016aen} 
  T.~Basile, X.~Bekaert and N.~Boulanger,
  ``Mixed-symmetry fields in de Sitter space: a group theoretical glance,''
  JHEP {\bf 1705}, 081 (2017)
  [arXiv:1612.08166 [hep-th]].
	
\bibitem{Metsaev:1995re} 
  R.~R.~Metsaev,
  ``Massless mixed symmetry bosonic free fields in d-dimensional anti-de Sitter space-time,''
  Phys.\ Lett.\ B {\bf 354}, 78 (1995).
	
\bibitem{Metsaev:1997hi} 
  R.~R.~Metsaev,
  ``Free totally (anti)symmetric massless fermionic fields in d-dimensional anti-de Sitter space,''
  Class.\ Quant.\ Grav.\  {\bf 14}, L115 (1997)
  [hep-th/9707066].
  ``Fermionic fields in the d-dimensional anti-de Sitter space-time,''
  Phys.\ Lett.\ B {\bf 419}, 49 (1998)
  [hep-th/9802097].
  
\bibitem{Evans}
  N.~T.~Evans, ``Discrete Series for the Universal Covering Group of the 3 + 2 de Sitter Group'', 
  J.\ Math.\ Phys.\ {\bf 8}, 170 (1967).
  
\bibitem{Gunaydin:1998jc} 
  M.~Gunaydin, D.~Minic and M.~Zagermann,
  ``Novel supermultiplets of SU(2,2|4) and the AdS(5) / CFT(4) duality,''
  Nucl.\ Phys.\ B {\bf 544}, 737 (1999)
  [hep-th/9810226].
  
\bibitem{Metsaev:1997nj} 
  R.~R.~Metsaev,
  ``Arbitrary spin massless bosonic fields in d-dimensional anti-de Sitter space,''
  Lect.\ Notes Phys.\  {\bf 524}, 331 (1999)
  [hep-th/9810231].

\bibitem{Deser:2001us} 
  S.~Deser and A.~Waldron,
  ``Partial masslessness of higher spins in (A)dS,''
  Nucl.\ Phys.\ B {\bf 607}, 577 (2001)
  [hep-th/0103198].

\bibitem{Boulanger:2008up} 
  N.~Boulanger, C.~Iazeolla and P.~Sundell,
  ``Unfolding Mixed-Symmetry Fields in AdS and the BMV Conjecture: I. General Formalism,''
  JHEP {\bf 0907}, 013 (2009)
  [arXiv:0812.3615 [hep-th]].

\bibitem{Skvortsov:2009zu} 
  E.~D.~Skvortsov,
  ``Gauge fields in (A)dS(d) and Connections of its symmetry algebra,''
  J.\ Phys.\ A {\bf 42}, 385401 (2009)
  [arXiv:0904.2919 [hep-th]].

\bibitem{Singh:1974qz} 
  L.~P.~S.~Singh and C.~R.~Hagen,
  ``Lagrangian formulation for arbitrary spin. 1. The boson case,''
  Phys.\ Rev.\ D {\bf 9}, 898 (1974).
  
\bibitem{Zinoviev:2009gh} 
  Y.~M.~Zinoviev,
  ``Towards frame-like gauge invariant formulation for massive mixed symmetry bosonic fields. II. General Young tableau with two rows,''
  Nucl.\ Phys.\ B {\bf 826}, 490 (2010)
  [arXiv:0907.2140 [hep-th]].

\bibitem{Beccaria:2015vaa} 
  M.~Beccaria and A.~A.~Tseytlin,
  ``On higher spin partition functions,''
  J.\ Phys.\ A {\bf 48}, no. 27, 275401 (2015)
  [arXiv:1503.08143 [hep-th]].
  
\bibitem{Giombi:2013fka} 
  S.~Giombi and I.~R.~Klebanov,
  ``One Loop Tests of Higher Spin AdS/CFT,''
  JHEP {\bf 1312}, 068 (2013)
  [arXiv:1308.2337 [hep-th]].

\bibitem{Giombi:2014iua} 
  S.~Giombi, I.~R.~Klebanov and B.~R.~Safdi,
  ``Higher Spin AdS$_{d+1}$/CFT$_d$ at One Loop,''
  Phys.\ Rev.\ D {\bf 89}, no. 8, 084004 (2014)
  [arXiv:1401.0825 [hep-th]].

\bibitem{Skvortsov:2017ldz} 
  E.~D.~Skvortsov and T.~Tran,
  ``AdS/CFT in Fractional Dimension and Higher Spin Gravity at One Loop,''
  arXiv:1707.00758 [hep-th].

\bibitem{unpubl}
  R.~R.~Metsaev, to appear.
  
  

\bibitem{Schuster:2014xja} 
  P.~Schuster and N.~Toro,
  ``A new class of particle in 2 + 1 dimensions,''
  Phys.\ Lett.\ B {\bf 743}, 224 (2015)
  [arXiv:1404.1076 [hep-th]].

\bibitem{Metsaev:unpublished}
R.~R. Metsaev, ``{Ph.D. Thesis}.'' 1991.


\bibitem{Metsaev:1991nb}
R.~R. Metsaev, ``{S matrix approach to massless higher spins theory. 2: The
  Case of internal symmetry},''
{{\em Mod. Phys. Lett.}
  {\bfseries A6} (1991) 2411--2421}.

\bibitem{Metsaev:1991mt} 
  R.~R.~Metsaev,
  ``Poincare invariant dynamics of massless higher spins: Fourth order analysis on mass shell,''
  Mod.\ Phys.\ Lett.\ A {\bf 6}, 359 (1991).
  
\bibitem{Ponomarev:2016lrm} 
  D.~Ponomarev and E.~D.~Skvortsov,
  ``Light-Front Higher-Spin Theories in Flat Space,''
  J.\ Phys.\ A {\bf 50}, no. 9, 095401 (2017)
  [arXiv:1609.04655 [hep-th]].

  
\bibitem{Roiban:2017iqg} 
  R.~Roiban and A.~A.~Tseytlin,
  ``On four-point interactions in massless higher spin theory in flat space,''
  JHEP {\bf 1704}, 139 (2017)
  [arXiv:1701.05773 [hep-th]].

\bibitem{Taronna:2011kt} 
  M.~Taronna,
  ``Higher-Spin Interactions: four-point functions and beyond,''
  JHEP {\bf 1204}, 029 (2012)
  doi:10.1007/JHEP04(2012)029
  [arXiv:1107.5843 [hep-th]].
\bibitem{Sleight:2016xqq} 
  C.~Sleight and M.~Taronna,
  ``Higher-Spin Algebras, Holography and Flat Space,''
  JHEP {\bf 1702}, 095 (2017)
  [arXiv:1609.00991 [hep-th]].
  
\bibitem{Taronna:2017wbx} 
  M.~Taronna,
  ``On the Non-Local Obstruction to Interacting Higher Spins in Flat Space,''
  JHEP {\bf 1705}, 026 (2017)
  [arXiv:1701.05772 [hep-th]].

\bibitem{Flato:1978qz} 
  M.~Flato and C.~Fronsdal,
  ``One Massless Particle Equals Two Dirac Singletons: Elementary Particles in a Curved Space. 6.,''
  Lett.\ Math.\ Phys.\  {\bf 2}, 421 (1978).

\bibitem{Zinoviev:2017rnj} 
  Y.~M.~Zinoviev,
  ``Infinite spin fields in d = 3 and beyond,''
  arXiv:1707.08832 [hep-th].

\bibitem{MN}
  M. Najafizadeh, ``Modified Wigner equations and continuous spin gauge field'' arXiv:1708.00827.
  
\end{thebibliography}
\end{document}